\documentclass[11pt]{article}

\usepackage{sectsty,amsmath}
\usepackage{graphicx}

\usepackage{bm}
\usepackage{subfigure}

\def\bbbr{{\rm I\kern-0.23em R}}
\def\bbbn{{\rm I\kern-0.23em N}}
\def\bbbz{{\mathchoice {\hbox{$\sf\textstyle Z\kern-0.4em Z$}}
        {\hbox{$\sf\textstyle Z\kern-0.4em Z$}}
        {\hbox{$\sf\scriptstyle Z\kern-0.3em Z$}}
        {\hbox{$\sf\scriptscriptstyle Z\kern-0.2em Z$}}}}
\def\bbbc{{\mathchoice {\setbox0=\hbox{$\displaystyle\rm C$}\hbox{\hbox
        to0pt{\kern0.4\wd0\vrule height0.95\ht0\hss}\box0}}
        {\setbox0=\hbox{$\textstyle\rm C$}\hbox{\hbox
        to0pt{\kern0.4\wd0\vrule height0.95\ht0\hss}\box0}}
        {\setbox0=\hbox{$\scriptstyle\rm C$}\hbox{\hbox
        to0pt{\kern0.4\wd0\vrule height0.95\ht0\hss}\box0}}
        {\setbox0=\hbox{$\scriptscriptstyle\rm C$}\hbox{\hbox
        to0pt{\kern0.4\wd0\vrule height0.95\ht0\hss}\box0}}}}

\newcommand{\Z}{\bbbz}

\newcommand{\beq}{\begin{equation}}
\newcommand{\eeq}{\end{equation}}

\def\OO{\mathcal{O}}

\newcommand{\pa}{\partial}

\title{Dynamics of Nonlinear Lattices}
\author{Christopher Chong and P.G. Kevrekidis}

\begin{document}
\maketitle	



\begin{abstract}
    In this topical review we explore the dynamics of nonlinear lattices with a particular focus to Fermi-Pasta-Ulam-Tsingou type models that arise in the study of elastic media and, more specifically, granular crystals.
    We first revisit the workhorse of such
    lattices, namely traveling waves, both from a continuum, but also from a genuinely discrete perspective, both without and with a linear force component (induced by the so-called precompression).
    We then extend considerations to time-periodic states, examining dark breather structures in homogeneous crystals,
    as well as bright breathers in diatomic lattices. The last pattern that we consider extensively is the dispersive shock wave arising in the context of
    suitable Riemann (step) initial data. We show how the use of continuum (KdV) and discrete (Toda) integrable approximations
    can be used to get a first quantitative handle of the relevant waveforms. 
    In all cases, theoretical analysis is accompanied by numerical computations and, where possible, by a recap and illustration of prototypical experimental
    results.
    We close the chapter by offering a number of ongoing and potential future directions and associated open problems in the field.
\end{abstract}

\vspace{-1.5cm}

\centerline{\textbf{}}

\vspace{.75cm}

\section{Introduction}

The theme of nonlinear dynamical lattices is a very multifaceted one
spanning a diverse range of disciplines within Mathematics,
Physics and Engineering. Even if one restricts oneself to
Hamiltonian or principally Hamiltonian lattices, the purview
of an associated review effort is far too wide, given their
diverse impact in different thematic areas. These extend
from
nonlinear optics, e.g., in the nonlinear
dynamics of optical waveguides~\cite{LEDERER20081},
to atomic Bose-Einstein condensates and the associated
realm of optical lattices~\cite{RevModPhys.78.179},
and from models of the DNA double strand~\cite{Peybi} 
to ones of electrical and mechanical elements~\cite{remoissenet}.
The nonlinear waveforms that emerge in such systems,
including traveling waves, kinks and breathers
have, in turn, been summarized in various 
reviews~\cite{willis,Flach2008,Aubry06}
and books~\cite{FPUreview,kev09,DPbook} (see also relevant
chapters in~\cite{floyd,p4book}).
Dispersive shock waves are, arguably, less 
extensively treated in the discrete realm~\cite{granularBook},
although they have been widely popular in their continuum form in dispersive systems~\cite{Mark2016}.

Within the confines of the present chapter, and given 
its specific parameters (including both its desired
length, but also its target audience), we have thus 
opted to focus our attention on a particular example
application and the associated developments over the 
past two decades, namely the theme of granular 
crystals~\cite{Nester2001,sen08,vakakis_review,yuli_book,granularBook}.
One of the motivations for the relevant choice stems from the
significant advances, including experimental ones,
within this field that have enabled the realization of
a wide palette of nonlinear excitations including
the main workhorse of relevant studies in the form
of the traveling wave~\cite{Nester_dimer,sen08,yuli_book},
the subsequently experimentally accessed dispersive
shock waves~\cite{granularBook}, but also the more recently
proposed and realized discrete breathers~\cite{granularBook},
as well as extensions of these structures in 
heterogeneous~\cite{yuli_book} and higher-dimensional
settings.

Another motivation for the choice of the granular
crystals as a ``paradigm of choice'' for the illustration
of the far-reaching impact of nonlinear dynamical lattices
is its substantial potential as a prototype for a wide
range of concepts of interest to practical applications.
For instance, such crystals have been proposed as a configuration
that can enable the manifestation of the concept of ``acoustic 
lenses'' that may localize the energy in the form of 
``sound bullets''~\cite{Spadoni}. Another intriguing direction
involves the use of such crystals (in the presence of
defects) for the realization of acoustic switching and
for inducing rectification in the work of~\cite{Nature11,theocharis_review}.
More recently, variants of such setups have been proposed
for the implementation of acoustic gates and logic elements~\cite{Li_switch}.
Yet another practical application has involved the use
of granular crystal sensors (via the scattering 
of traveling waves) to assess whether bones are healthy
or osteoporotic~\cite{bone}.

Our presentation of some of the principal ideas and nonlinear patterns of
such metamaterial lattices will be structured
as follows. In section~\ref{TW}, we will present an overview
of the traveling waves in such systems. This will be followed
by an exposition of the conditions under which time-periodic,
exponentially-localized waveforms (so-called discrete breathers)
can arise in~\ref{DB}. The more recent, somewhat more exotic
and still rather elusive description of the dispersive shock
waves will be explored in~\ref{DSW}. We end the chapter summarizing and presenting a number of current and future
directions in~\ref{SFD}.

\section{Traveling Waves}
\label{TW}

A seminal point towards the developments of interest to granular crystals
(hereafter abbreviated as GCs) stemmed from the work of H. Hertz in 1881~\cite{Hertz}
who characterized the force between two elastic bodies under what is now known
as Hertz's law:
\begin{equation}\label{hertz}
	F(\delta) = A [\delta]_+^{3/2}\,, \quad A = \frac{E\sqrt{2R}}{3(1-\nu^2)}\,,
\end{equation}	
Here, $E$ stands for  the elastic (Young's) modulus and characterizes the material
stiffness, while $\nu$ represents the Poisson ratio, i.e., the tendency for
transverse expansion under longitudinal compression; $R$ stands for the body's
radius of curvature at contact. Here, we will mostly deal with spheres, hence it
corresponds to the sphere's radius. The $+$ stands to indicate that the force is
only active under contact (compression). The exponent $3/2$, i.e., the genuinely
nonlinear character of the associated force is a principal source of the appeal
of the relevant law towards the emergence of nonlinear excitations and for our
considerations herein we will limit ourselves to that case. Yet, it is relevant
to mention in passing that in other settings such as O-rings, cylindrical particles
or hollow spheres the nature of the force may vary~\cite{Johnson}.

Now assembling a lattice of such spherical beads in the form of a GC
(or a nonlinear mechanical metamaterial, as it is often referred to) leads
to equations of motion encompassing the interaction of each bead with its nearest
neighbors in the form of:
 \begin{equation}
 	\ddot{u}_n = \frac{A_n}{M_n}[\delta_{0,n}+u_{n-1}-u_n]_+^{p}-
 	\frac{A_{n+1}}{M_n}[\delta_{0,n+1}+u_n-u_{n+1}]_+^{p}.
 \label{eq:model}
 \end{equation}
Here, $u_n$ stands for the displacement of the $n$-th particle of mass $M_n$ (measured from the
equilibrium). The so-called static precompression displacement (or simply precompression)
is given by $\delta_n = ({F_0}/{A_n})^{1/p}$ and is associated with the (pre)compression force
$F_0$. The expression for $A_n$ is more complex for non-identical 
neighbors~\cite{Nester2001,sen08}, 
\begin{equation} \label{material}
A_{n} = \frac{4 E_{n}E_{n+1}\sqrt{\frac{R_{n}R_{n+1}}{\left(R_{n}+R_{n+1}\right)}}}%
{3E_{n+1}\left(1-\nu_{n}^2\right) + 3 E_{n}\left(1-\nu_{n+1}^2\right)  },
\end{equation}
with the indices representing corresponding neighboring beads. Yet, here, for
much of our discussion we will restrict considerations to homogeneous chains. 

The strain formulation, where $y_n = u_{n+1} - u_n$, is also useful.
In this case
the original equation of motion Eq.~\eqref{eq:model}  becomes:
\begin{equation}
	\ddot{y}_n =  2 F(y_n) - F(y_{n-1}) + F(y_{n-1}) 
\label{strain}
\end{equation}
where $F(y) = [\delta_0 - y_{n}]_+^{3/2}$ and for simplicity
we have assumed a time rescaling absorbing the factor of $A/M$.

When the precompression is finite, and indeed much larger than the 
pairwise relative displacements of the beads $\delta_0 \gg |y_n|$, it is
straightforward to observe that the model Eq.~(\ref{eq:model}) can be Taylor
expanded and yields to leading order a discrete Laplacian, along with quadratic
and subsequently cubic corrections similar to those emerging in the 
$\alpha$ and $\beta$, respectively, installments of the famous Fermi-Pasta-Ulam
model~\cite{FPU55,FPUreview,zabu} that has been recently renamed to 
Fermi-Pasta-Ulam-Tsingou (FPUT)~\cite{daux} to acknowledge
contributions of Tsingou, a convention that we will use hereafter. 
In terms of the strain variable, the FPUT model is simply Eq.~\eqref{strain}
with $F(y) = K_2 y + K_3 y^2 + K_4 y^3$, where the $K_j$ are coefficients,
stemming, e.g., from a Taylor expansion.
As such, it has been rigorously established that the model can be connected
to (i.e., approximated by) the famous Korteweg-de Vries (KdV) 
\begin{equation}
\partial_{T}  Y +  \partial_X^3 Y + Y \partial_X Y = 0.     \label{kdv}
\end{equation}
using the ansatz $y_n(t) = -\epsilon^2 \sigma^2/ (24 K_3) Y(X,T)$, where
$X = \epsilon( n - \sigma t)$, $T = \sigma \epsilon^3 t/24$ and where
$\sigma=\sqrt{A p \delta_0^{p-1}/M}$ is the sonic
speed limit for homogeneous chains.
Then, the weakly supersonic waves of the FPUT (and accordingly of the GC) can
be approximated by the continuum solitons of the KdV equation
and indeed they inherit not only the existence properties, but also the
stability properties of such waves~\cite{pegof2,pegof3,pegof4}. This perspective
has been leveraged to connect the GC traveling waves (TWs) with those of the
KdV, including as regards the famous  collision example of
the tall and thin and fast soliton overtaking the short and fat and slow one~\cite{Shen}.
Indeed, in this work, connections along the same vein were also forged
with another famous integrable, yet genuinely discrete model, the Toda lattice.
In terms of the strain, it is given by Eq.~\eqref{strain} with
$F(y) = e^y$. The connection is made by simply matching the linear and quadratic
Taylor coefficients of the granular chain and Toda lattice.
The Toda reduction allows one to describe TWs and their interactions not only in the unidirectional
case (as the KdV analysis allows), but also in the bidirectional setting of
waves colliding when moving in opposite directions. 

Our focus in this section will be primarily on the more mathematically interesting case
where the precompression is absent, i.e., $\delta_0=0$. For a considerable while,
it was believed that this setting leads to, arguably, the most physically relevant
example of a structure that was proposed in the 1990's in the context of nonlinear
wave models devoid of linear dispersion (and, thus, featuring only nonlinear dispersion),
the so-called compactons~\cite{Rosenau93}. 
This
belief stemmed from the attempt to produce a quasi-continuum model description of
the underlying discrete system of Eq.~\eqref{eq:model}. Absorbing the factor of $A/M$
through a rescaling and using a(n artificially introduced) slow scale of space and
time $u_n(t) = U(X,T)$, where $X = \epsilon n$, $T = \epsilon t$, Nesterenko and
collaborators obtained the long wavelength model of the form:
\begin{eqnarray}
  \pa_T^2 U= -\pa_X \left[  (-\pa_X U)^{p} + \frac{\epsilon^2}{12} \left( \pa_X^2 (-\pa_X U)^{p}  - \frac{p(p-1)}{2} (-\pa_X U)^{p-2}   \pa_X^2 U^2\right)
    \right]. 
\label{ch3_tw2}
\end{eqnarray}
One can also derive the corresponding equation for the strain defined as $Y=\pa_X U$.
The key realization of Nesterenko and coworkers~\cite{Nester2001,Coste} was that this PDE
in the so-called co-traveling frame associated with the variable  $\xi=X-c T$ can yield
traveling waveforms as:
 \begin{eqnarray}
    Y(X-cT)&&=c^m A \cos^{m}(B (X-cT)), \\
    &&\quad m=2/(p-1), \quad B=\sqrt{6 (p-1)^2/(p (p+1))}.
    \label{ch3_tw4}
  \end{eqnarray}
Notice that these are periodic solutions, hence the proposed rubric for 
constructing the compactons, as is typically the case in such settings,
is to isolate a period of this trigonometric solution and glue it with 0's
on its two sides to construct such a pattern. 

An alternative approach brought forth more recently~\cite{pikovsky} was to consider the strain
variables already at the discrete level, see Eq.~\eqref{strain}.
 Importantly,
this is a Hamiltonian system whose Hamiltonian function can be written as:
$$H = \sum_{n\in\Z} \frac{1}{2} M_n \dot{u}_n^2 + V_n(u_{n+1} - u_n)$$
while the corresponding potential reads:
\begin{equation}
V_n(y_n) =  \frac{2}{5} A_{n+1}[ \delta_{0,n} - y_n]^{5/2} + \phi_n,  \qquad  \phi_n =  - \frac{2}{5} A_{n+1}[ \delta_{0,n} ]^{5/2} - A_{n+1}[ \delta_{0,n} ]^{3/2} y_n       \label{pot}
\end{equation}
It is notable that this approach is useful even in the presence of precompression
because as observed in~\cite{MacKay99}, it formulates the problem with a potential
featuring $V_n'(0) = 0$ and $V_n''(0) \geq 0$, thus enabling the use of the classic
result of~\cite{Wattis} towards a variational, yet rigorous proof of existence of a traveling wave. 

\begin{figure}[ht]
 \begin{minipage}{0.45\textwidth}
    \includegraphics[width=2in]{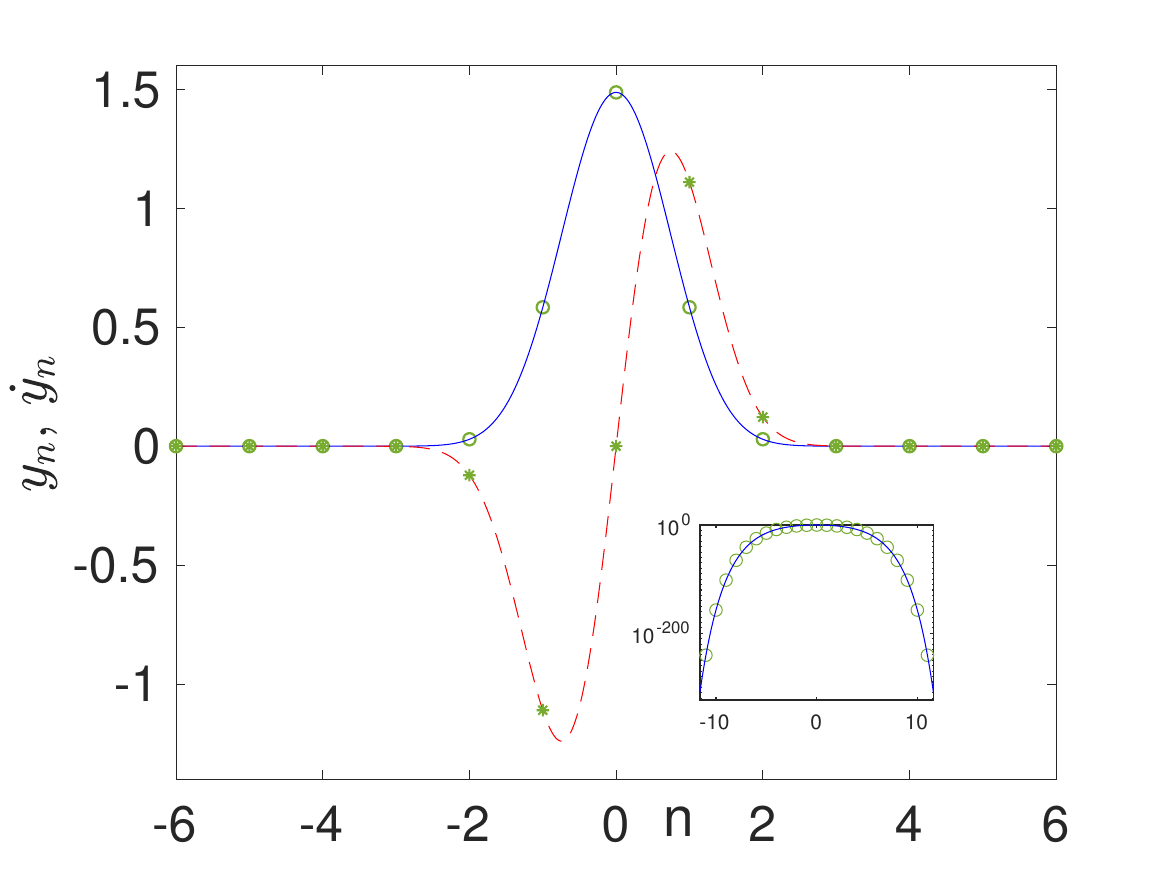}
    \caption{Traveling wave theoretical profile (blue solid) and its 
    derivative (green dashed). The decay is shown in semilog in the inset.
    Lattice nodes denoted by the green points}\label{fig1.2}
\end{minipage}
\hfill 
 \begin{minipage}{0.45\textwidth}
    \includegraphics[width=2in]{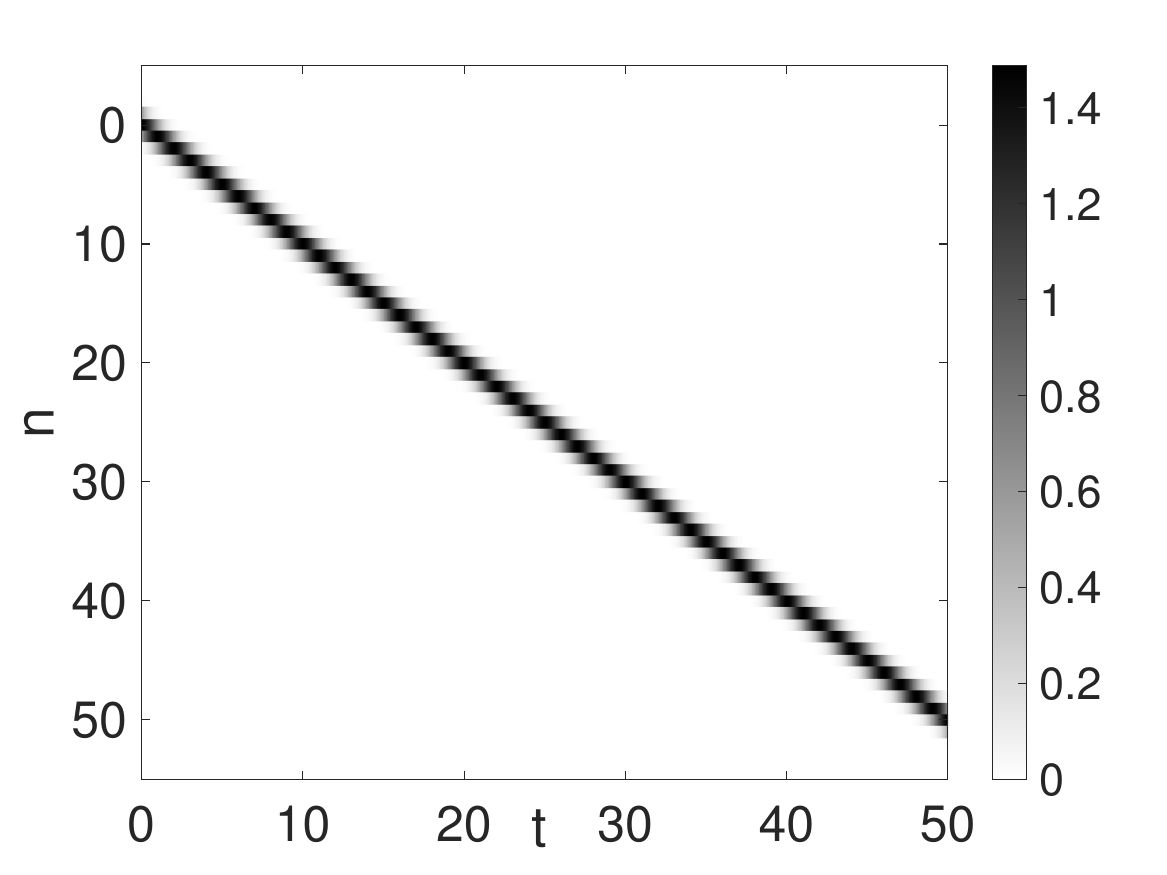}
    \caption{Space-time evolution of the exact traveling wave strain solution for
    a speed of $c=1$.} \label{fig1.3}
\end{minipage}

 \begin{minipage}{0.45\textwidth}
    \includegraphics[width=2in]{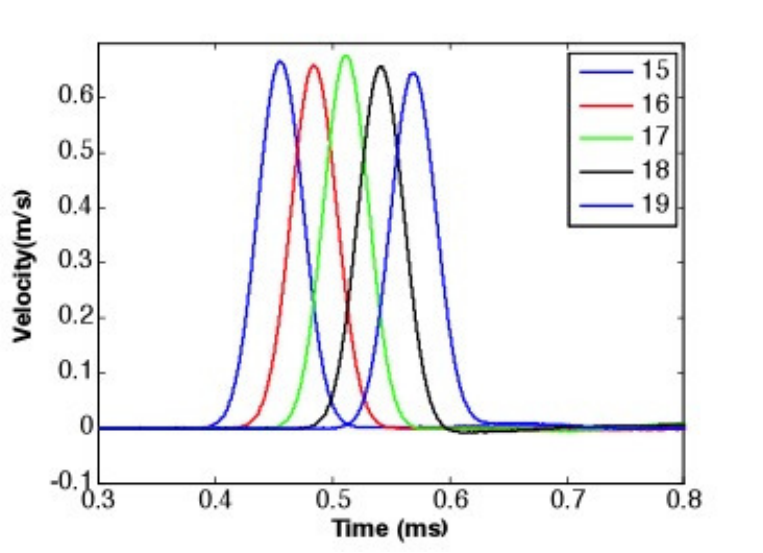}
    \caption{Experimental measurement of the velocity in adjacent nodes showcasing the traveling wave
    nature of the pulse in a granular crystal.Figure from Ref. \cite{JKdimer}. Used with permission.} 
    \label{fig1.4}
    \end{minipage}
    \hfill 
 \begin{minipage}{0.45\textwidth}
    \includegraphics[width=2in]{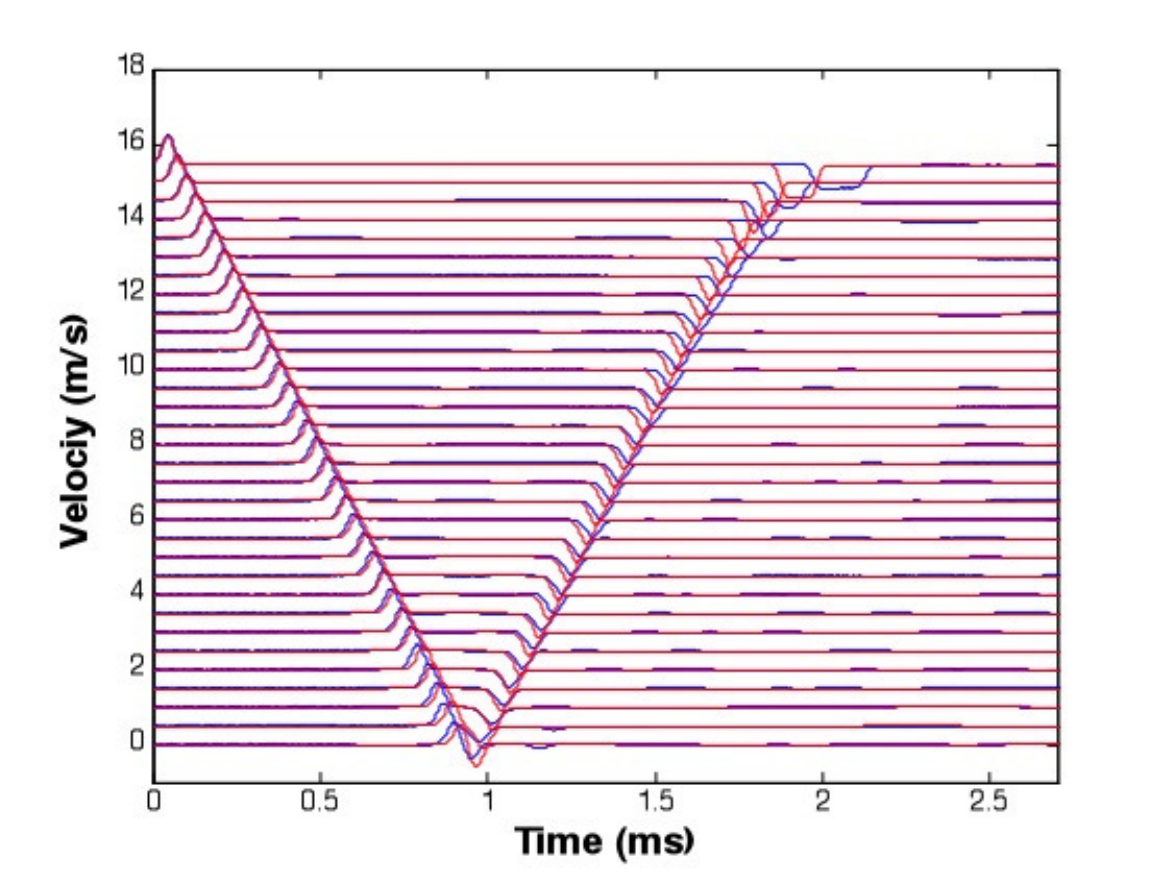}
    \caption{Collecting experimental measurements of the left panel into a space-time plot illustrating the
    traveling wave propagation along the granular crystal.
    Figure from Ref. \cite{JKdimer}. Used with permission.} \label{fig1.5}
\end{minipage}
\end{figure}

The approach of~\cite{pikovsky} was to then retrieve a quasi-continuum model directly
at the level of the strains (without passing from a quasi-continuum of displacements as before), 
in which case the resulting PDE reads:
\begin{eqnarray}
	\pa_{T}^2Y=\pa_X^2(Y^p) + \frac{\epsilon^2}{12} \pa_X^4(Y^p).
\label{ch3_tw7}
\end{eqnarray}
Remarkably the latter features a compacton too:
\begin{eqnarray}
  Y(X-cT)=&&c^m A \cos^{m}(\tilde{B} (X-cT)), \\
  &&m=2/(p-1), \quad \tilde{B}=\sqrt{3} (p-1)/p.
  \label{ch3_tw8}
\end{eqnarray}
Yet, equally remarkably, as is already evident by comparing Eq.~(\ref{ch3_tw7}) with
Eq.~(\ref{ch3_tw2}), the two procedures [quasi-continuum limit derivation 
and strain variable transformation]
do {\it not} commute, given the difference between the resulting equations and
the associated compactons.

A natural question is which of the two functional waveforms is the ``right'' one
in describing the actual traveling wave. While both of them are proximal to the
TW, especially so for $p=3/2$, the actual answer is, surprisingly, {\it neither}. 
Indeed, the relevant decay of the wave's tail is {\it doubly exponential} in nature. 
This was originally argued in~\cite{Chatterjee}, but was rigorously shown for this 
problem in the work of~\cite{pego1}. This was a byproduct of a reformulation of the
relevant traveling wave problem (which is tantamount to an advance-delay
differential equation) in Fourier space. The latter idea dates back to the work
of~\cite{Hochstrasse}. More specifically, seeking a
traveling wave in the form $y_n(t) = \Phi(n - ct,t) = \Phi(\xi,t)$ yields the problem:
\begin{multline}\label{step_2}
 	\partial_t^2{\Phi}(\xi,t)=-c^{2} \partial_\xi^2{\Phi}(\xi,t)+2c%
 \partial_{\xi \,t }{\Phi}(\xi,t)+  \\ \Big\lbrace\left( {\Phi}(\xi-1,t)\right)_{+}^{3/2}-%
2\left( {\Phi}(\xi,t)\right)_{+}^{3/2}+\left( {\Phi}(\xi+1,t)\right)_{+}^{3/2}\Big\rbrace\,, 
\end{multline}
which, subsequently, for time-independent solutions in the co-traveling frame (i.e., only
dependent on $\xi$) and through the Fourier transform $\phi(\xi)=\int_{-\infty}^{\infty} \hat{\phi}(k) e^{i k \xi} dk$
results in the fixed point scheme:
\begin{equation} \label{fft}
  \hat{\phi}(k) = \frac{1}{c^2} {\rm sinc}^2\left(\frac{k}{2}\right)\widehat{\phi^{3/2}}\,.
\end{equation}
One can then Fourier transform back into real space to obtain the fixed point scheme in the
real space variable in the form:
\begin{equation} \label{fp}
	\phi(\xi) = \Lambda \ast \phi^{3/2} = \int_{-\infty}^{\infty} \Lambda(\xi-y) \phi^{3/2}(y) dy\,
\end{equation}
The relevant kernel is $\Lambda(\xi)= (1/c^2) [ \, 1-|\xi| \, ]_+$ and is referred to as the
tent function. This formulation enables the numerically exact 
(up to a prescribed numerical accuracy) computation of the relevant waveform
as is shown in Fig.~\ref{fig1.2}-\ref{fig1.3} [intended to also contain a comparison between the numerical TW, and the
two analytical ones of~\cite{Nester2001} and~\cite{pikovsky}]; this was illustrated
since the early works of~\cite{Hochstrasse} in FPUT problems and more concretely for
the granular problem in~\cite{pego1}. The latter work also showcased the decay of
the strain pulse associated with the TW. Assuming $c=1$, one can change to the variable
 $z=\xi-y$ to rewrite Eq.~(\ref{fp}) as:
 \begin{equation} \label{eqn23}
	\phi(\xi+1) =\int_{-1}^1 \Lambda(z) \phi^{3/2}(\xi+1-z) dz 
	\leq 
	\phi^{3/2}(\xi) \Rightarrow \phi(\xi+n) \leq \phi(\xi)^{(\frac{3}{2})^n}.
\end{equation}
Here, the unit integral of the tent function has been used, as well as 
the positive, monotonically decreasing tail of the TW.

In the seminal works of pioneers in the field 
summarized, e.g., in~\cite{Nester2001} and~\cite{sen08} it was only possible to measure the speed of individual
beads and to attempt to reconstruct, based on that, features such as the speed of the traveling wave. 
Yet, in recent times, remarkable experimental techniques such as Laser Doppler Vibrometry (LDV) have
ushered the study of GCs into a new experimental era whereby it is possible to scope each 
bead within the configuration and use the Doppler effect to infer its speed as a function of time. A systematic repetition of the experiment and scoping
of each bead enables the identification not only of the traveling nature of the structure by measuring
some of the beads (see, e.g., Fig.~\ref{fig1.4}), but also the visualization of the full evolution 
of the wave, as shown in Fig.~\ref{fig1.5}.

\begin{figure}[h]
\ \begin{minipage}{0.45\textwidth}
    \includegraphics[width=2.1in]{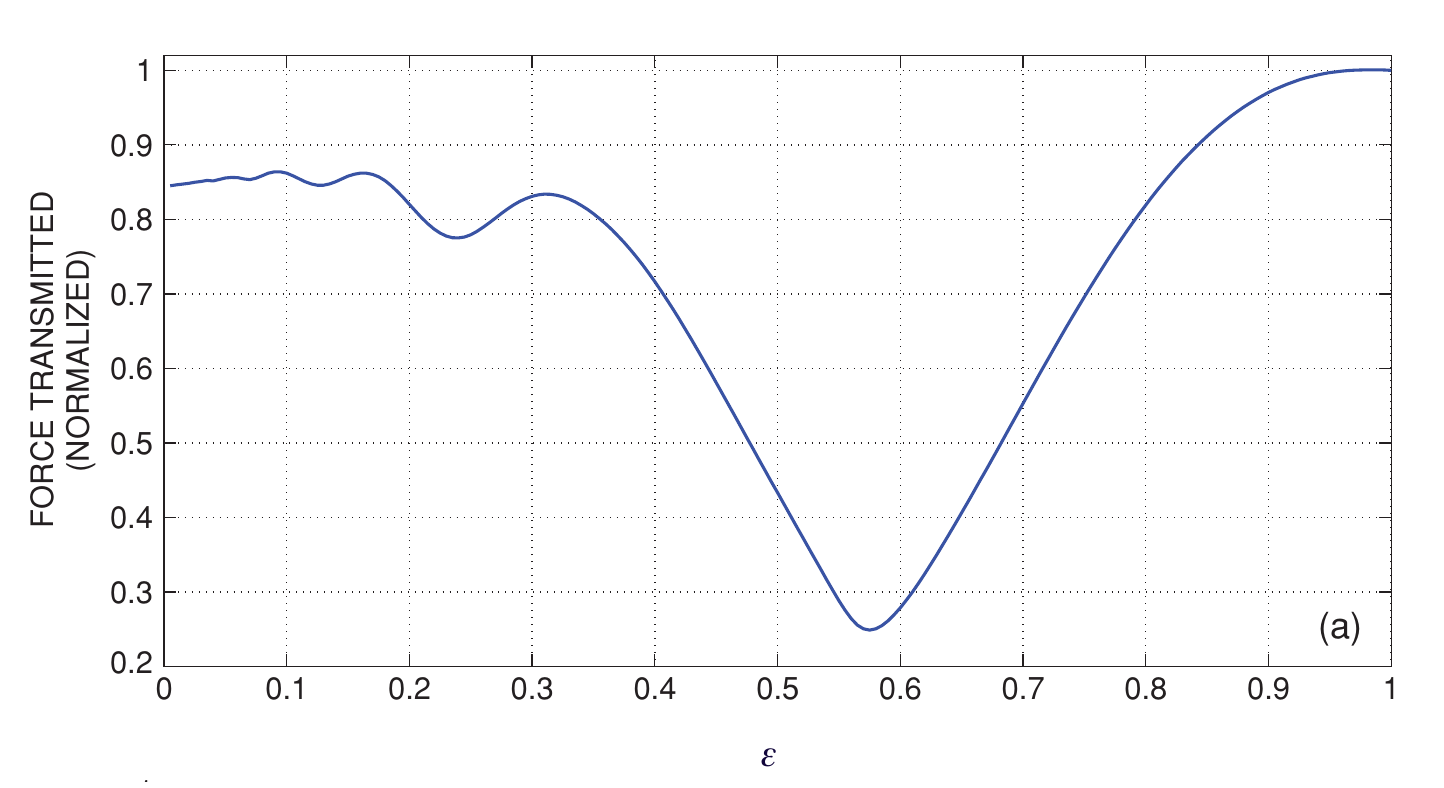}
    \caption{The numerical experiment showing
    the (normalized) response of a dimer chain, in terms of a transmitted force, as a function
    of the mass ratio $\epsilon$.
    Figure from Ref. \cite{Jayaprakash1}. Used with permission.
    }\label{fig2.2}
\end{minipage}
\hfill
\begin{minipage}{0.45\textwidth}
    \includegraphics[width=2in]{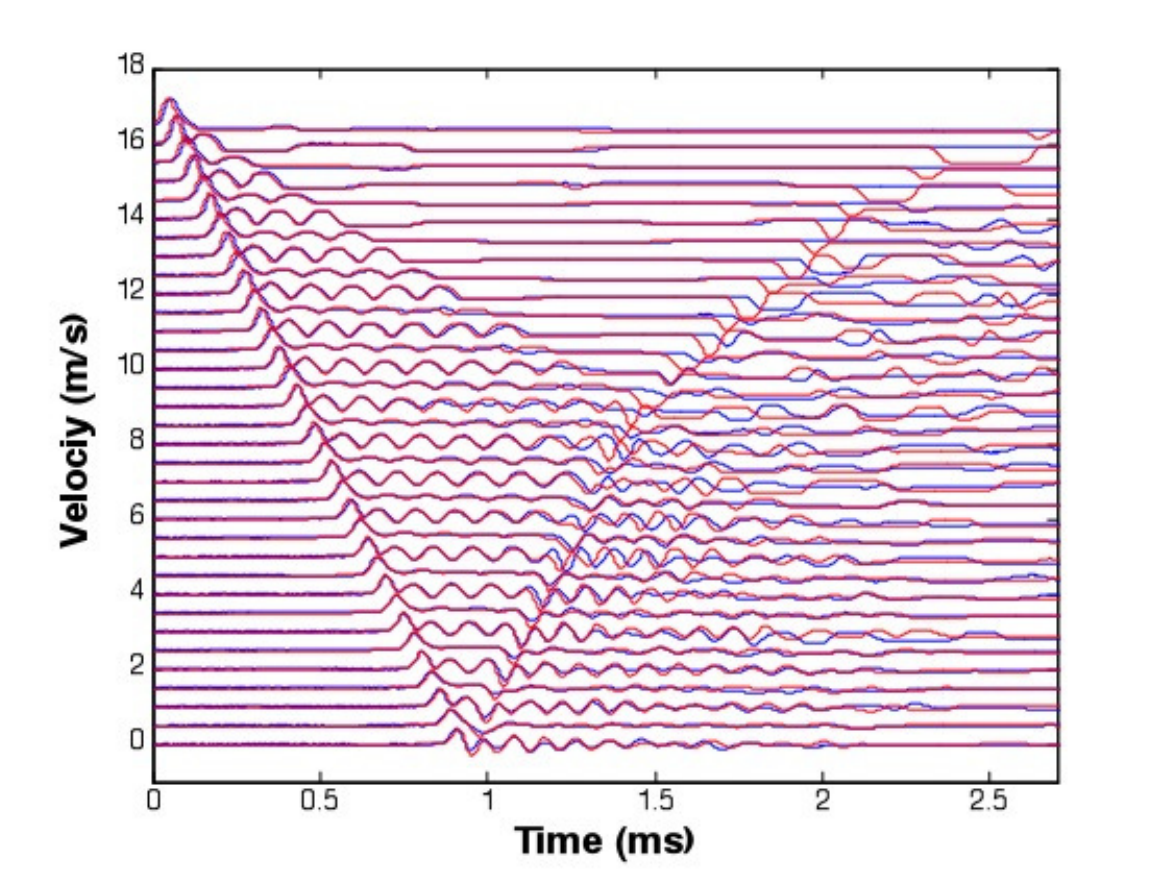}
    \caption{Space-time experimental evolution propagation associated with the case of $\epsilon=0.59$ showcasing
    the first and most dramatic resonance in a dimer chain 
    Figure from Ref. \cite{JKdimer}. Used with permission.
    }\label{fig2.3}
\end{minipage}
 \begin{minipage}{0.45\textwidth}
    \includegraphics[width=2in]{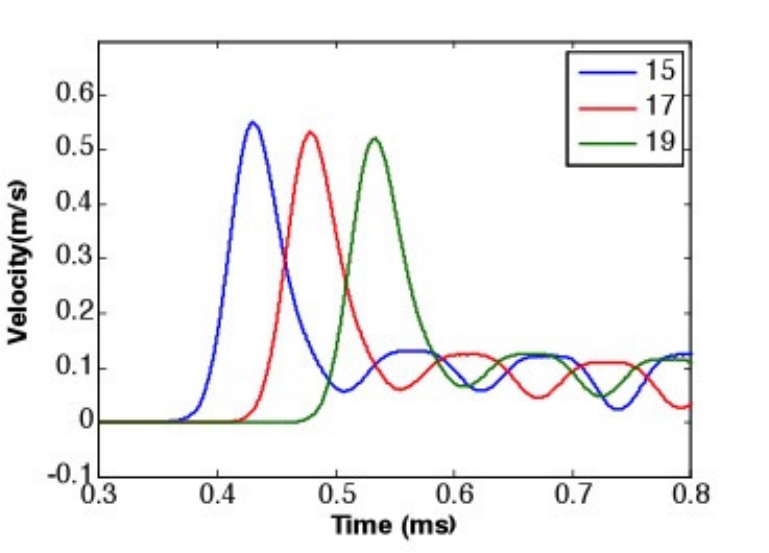}
    \caption{The evolution of the large mass beads of the top right panel case of $\epsilon=0.59$ as measured
    experimentally.
    Figure from Ref. \cite{JKdimer}. Used with permission.}\label{fig2.4}
\end{minipage}
\hfill
 \begin{minipage}{0.45\textwidth}
    \includegraphics[width=2in]{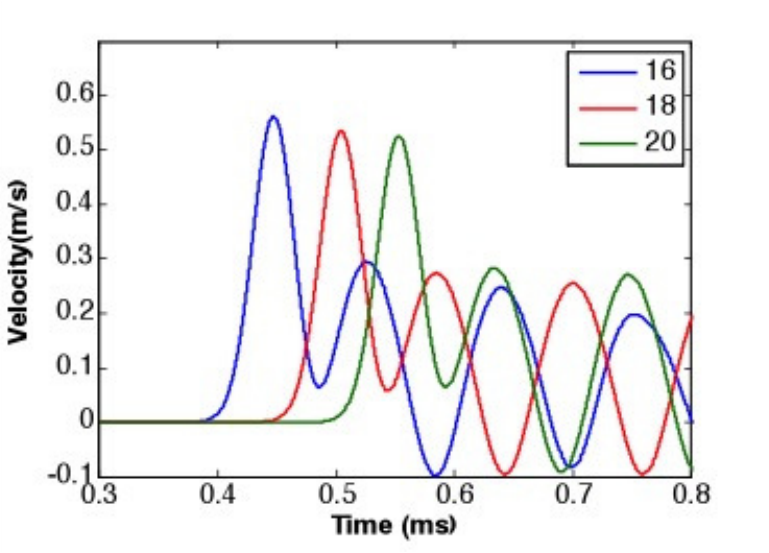}
    \caption{The evolution of the small mass beads of the top right panel case of $\epsilon=0.59$ as measured
    experimentally.
    Figure from Ref. \cite{JKdimer}. Used with permission.}\label{fig2.5}
\end{minipage}
\end{figure}

All of the above developments already introduced a mathematically, physically and experimentally
challenging problem at the level of the monomer GC. Interestingly, however, the story gets far more
complex once one considers heterogeneous chains. The study of such chains received considerable attention (including
through experimental studies) in works after the mid-2000s; see, e.g.,~\cite{mason1,mason2}. 
These studies involved the prototypical realm of dimers (i.e., granular chains consisting of two types
of beads, e.g., consisting of two different materials, or of the same material but different sizes),
as well as trimers (3 different types of beads) or more complex heterogeneous chains of, e.g., 5:1 chains
consisting of five beads of one type and then one of another.
However, it was
not until the work~\cite{Jayaprakash1} (see also the later one of~\cite{Jayaprakash2}) that the truly
fundamental alterations to the homogeneous, well-defined TW propagation were realized. This realization
was subsequently extended to more complex chains in works such as those of~\cite{vak_star3,staros2}.
For a summary of the relevant findings, the interested reader is pointed to the authoritative
description of~\cite{yuli_book}.

Here we outline the most prototypical of these effects, namely the existence of so-called
resonances and anti-resonances in dimer granular chains~\cite{Jayaprakash1}. The originally
proposed Gedankenexperiment was as follows: we know that in a GC, if we excite a bead on one
end, in the absence of precompression, there are no sound waves, so the energy will be channeled
into (typically) a TW which will be found to unobstructedly propagate throughout the lattice,
as discussed above. What if we repeat this type of excitation in a dimer chain? The results were
quite {\it dramatic}. It was found that if the ratio of the two masses within a dimer were $0.59$,
not only was the initial energy not going to be fully transmitted, but, in fact, only {\it less than 30\%}
of it would make it through. This is showcased quite strikingly in the experimental
realization of the relevant setting that followed a few years later in the work of~\cite{JKdimer},
showcased in the space-time strain evolution of Fig.~\ref{fig2.4}. The theoretical transmission 
fractions predicted via numerical simulations can be seen (in comparison to the normalized
monomer case of $\epsilon=1$ ($\epsilon$ represents the ratio of the smaller to the larger mass))
in the panel of Fig.~\ref{fig2.2}. It is clear in Fig.~\ref{fig2.3} that both the larger masses
(whose speeds are represented in Fig.~\ref{fig2.4}, as well as the more intensely vibrating
smaller ones of Fig.~\ref{fig2.5} are shedding away energy as the wave passes through, leading
eventually (as shown in Fig.~\ref{fig2.3}) to the complete destruction of the wave. The latter
is caused through the interference of the wave with the radiative tails it sheds. Entirely contrary
to this resonance phenomenology is the case of $\epsilon=0.3428$, the (2nd after that of the monomer)
so-called anti-resonance whereby to every vibration of the larger mass correspond two vibrations of
the smaller one leading to a(n essentially) perfect transmission and a genuine traveling wave being
restored. The sequence of ``quanta'' of resonances and anti-resonances (in each of the latter, the
smaller mass performs one additional vibration per one vibration of the large mass) was elucidated
in the work of~\cite{Jayaprakash1} (and subsequently considered in other settings in the works mentioned
above) and was experimentally visualized in the work of~\cite{JKdimer}. In our view, it yields a 
remarkable illustration of how dramatically different the phenomenology of heterogeneous chains
can be, in comparison to homogeneous ones. Moreover, these considerations initiated a considerable
effort to understand the nature of waves in such diatomic lattices,  e.g., of the Toda or of the FPUT type,
from a more systematic mathematical perspective, respectively in works such as those of~\cite{perline}
and~\cite{faver}.

\section{Discrete Breathers}
\label{DB}

A breather is defined as a structure that is periodic in time and localized in space.
Breathers have a rich history, dating back to their first study
in the context of the Sine-Gordon (sG) equation; see, e.g.,~\cite{morris,floyd} for relevant reviews.
Breathers in space-time continuous systems, such as the sG model,
are somewhat special. While a breather's frequency may lie in a frequency
gap, coupling of higher order harmonics to the
continuous spectrum of the system via nonlinearity 
is generally unavoidable since the linear passband extends to infinity.
Such harmonic-coupling mechanisms to spatially extended plane waves
will be detrimental to the spatial localization, and ultimately
prevent the existence of genuine breather solutions.
The sG model is an exception to this, and the coupling to higher order harmonics is only avoided 
due to the integrablity of the equation. In contrast to spatially continuous systems, the passbands
of spatially discrete systems are bounded from above, and thus such coupling mechanisms can controllably be avoided~\cite{ST,page}. 
Hence, the existence of breathers in discrete systems
should be more generic. Indeed, the work
of~\cite{macaub} provided a mathematical foundation for this
expectation illustrating that under (fairly generic) non-resonance
conditions, discrete breathers should be expected to persist
in a large class of nonlinear, spatially extended discrete dynamical systems.
There has been a vast amount of research on discrete breathers,
see for example ~\cite{Flach2008}.

\begin{figure}[h]
\begin{minipage}{0.45\textwidth}
    \includegraphics[width=2in]{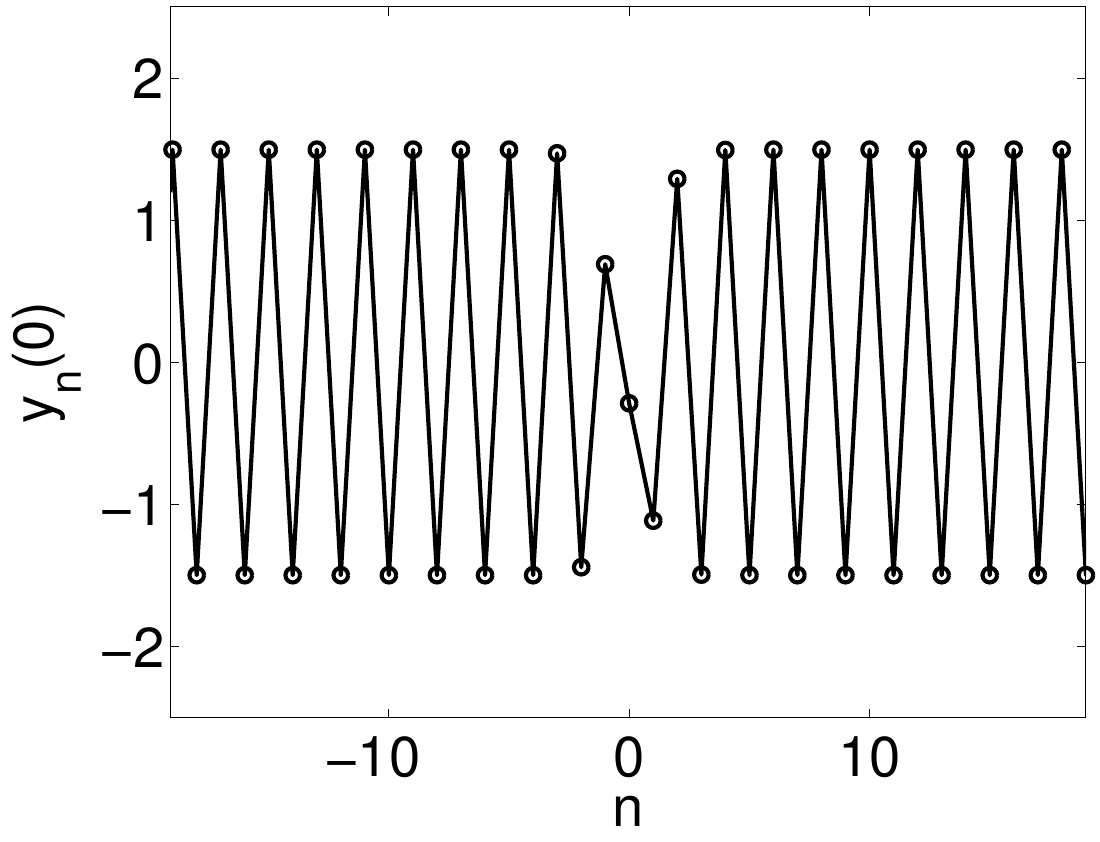}
    \caption{NLS approximation of a dark breather of the granular chain (markers). Solid lines are shown as a visual aid. The strain profile is shown.
    Figure from Ref. \cite{dark}. Used with permission.
    }\label{breather.1}
    \end{minipage}
    \hfill
\begin{minipage}{0.45\textwidth}
    \includegraphics[width=2in]{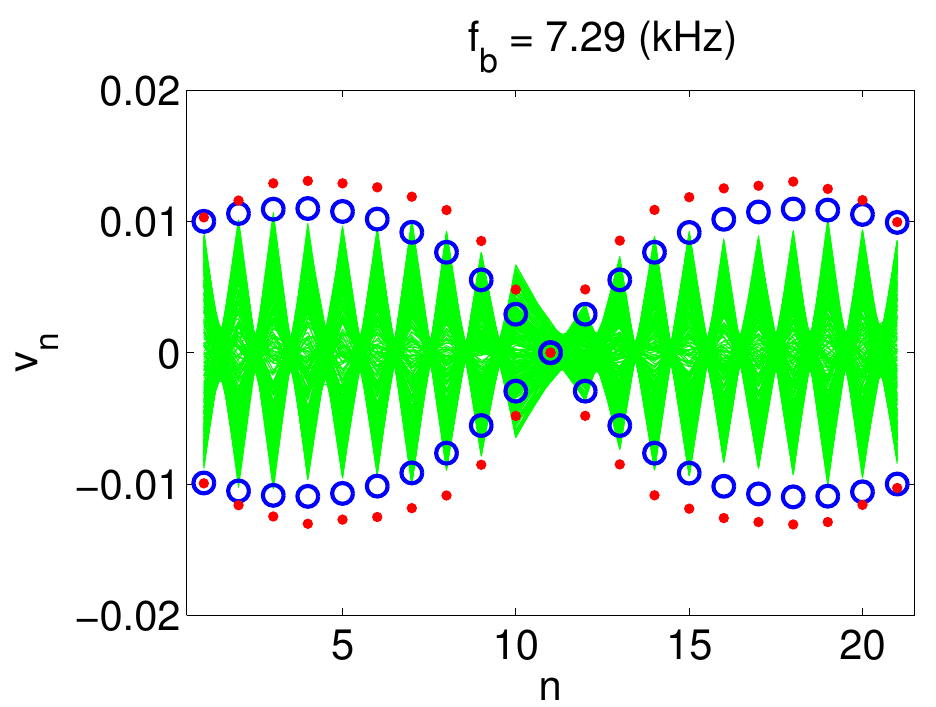}
    \caption{Experimental realization of a dark breather of the granular chain. Green line shows the time dynamics and the blue circles are the maxima/minima. The red dots correspond the model prediction. The velocity profile is shown. 
    Figure from Ref. \cite{dark2}. Used with permission.
    }\label{breather.2}
\end{minipage}
\end{figure}

In this section, we will focus on breathing structures of the granular chain.
What we will see is that bright discrete breathers 
(i.e., ones on top of a vanishing background)
are only possible in
non-homogeneous granular chains.  A related
structure, called a dark-breather, can be found in homogeneous granular chains,
which is where we start. We once again employ a multi-scale
analysis, this time using the ansatz
\begin{equation} \label{eq:NLSansatz}
y_n(t) = \epsilon Y(X,T) e^{i( k n + \omega t)}  + \mathrm{c.c.}, 
\end{equation}
where c.c. is the complex conjugate and the slow 
variables are $X=\epsilon ( n - c t)$ and $T = \epsilon^2 t$. This
ansatz is similar to the one we used to derive the KdV equation,
except now it includes a breathing
in time (with frequency $\omega$) and a spatial modulation with
wavenumber $k$. The amplitude and slow time have also been suitably modified. Substituting Eq.~\eqref{eq:NLSansatz} into Eq.~\eqref{strain}
leads to a hierarchy of equations. At $\OO(\epsilon)$, we
retrieve the dispersion relation of the system, i.e.,
$\omega^2=4 K_2 \sin^2 (\frac{k}{2})$. At the second order $\OO(\epsilon^2)$,
we are able to determine the group velocity $c$ to be
$c = -\omega'(k)$, where $\omega' = \frac{d\omega}{dk}$.
Subsequently, at the
cubic order $\OO(\epsilon^3)$, the Nonlinear Schr\"odinger (NLS) equation is retrieved in the form:
\begin{equation}\label{NLS}
 i \partial_T Y(X,T) + \nu_2 \partial_X^2 Y(X,T) + \nu_3 Y(X,T)|Y(X,T)|^2 = 0.
\end{equation}
See \cite{huang1,granularBook} for the details of the full derivation.
The coefficient of the dispersion reflects the (opposite of the) concavity
of the dispersion relation, i.e., $\nu_2 = -\omega''(k)/2$, while the
nonlinearity coefficient is given by a considerably more elaborate
expression. When the wavenumber is at the band edge ($k=\pi$), it simplifies to
$\left.\nu_3\right|_{k = \pi} = K_2^{-3/2} (3 K_2 K_4 - 4K_3^2) = B$.
At this wavenumber, the group velocity vanishes, and hence
the breathers will be standing. At this same band edge,
the concave down form of the dispersion relation gives
$\nu_2 >0$. However, for Taylor coefficients stemming
from the granular chain, we have that $B<0$, and hence
we are in the realm of the so-called defocusing NLS equation~\cite{DarkBook}
in which case one has the so-called dark soliton of the form,
\begin{equation}
	 Y(X,T) = \sqrt{\kappa/B} \tanh\left( \sqrt{ \frac{-\kappa}{2 \nu_2 }}(X - X_0) \right) e^{ i\kappa T}\,, \qquad \kappa <0 \,, 
\end{equation}
Thus, 
the corresponding reconstruction of the solution with $k=\pi$ of
the original granular chain is of the form:
\begin{eqnarray}\label{approx_james}
y_n(t) = 2 \epsilon (-1)^n   \sqrt{\frac{\kappa}{B}} \tanh \left( \sqrt{ \frac{-\kappa}{ 2 \nu_2 }}   \epsilon (n - x_0)   \right) \cos( \omega_b t ),   
\end{eqnarray}
where $\omega_b= \omega_0 + \kappa \epsilon^2$ is the breather frequency, with $\omega_0 = \omega(\pi)$
being the linear cutoff frequency (in this case, the top
of the acoustic branch of the dispersion relation). 
An example plot of this solution
is shown in Fig.~\ref{breather.1}. Notice that the solution has a non-zero background,
and has a density dip at the center of the lattice. This is the so-called dark breather, i.e., an exponentially localized,
temporally periodic dark pattern (on top of a background),
that additionally bears a staggered (alternating)
form herein due to the $k=\pi$ wavenumber.

Formula Eq.~\eqref{approx_james} provides an analytical approximation of the
dark breathers of the granular chain that is accurate under the assumptions
of which the formula was derived, namely with small amplitude, long wavelength,
and with frequency close to the cutoff. To explore dark breathers beyond these
assumptions, numerical computation is a useful tool.

We now provide a brief summary of how to compute time-periodic solutions in 
granular crystals, or lattices in general. More details can be found in \cite{Flach2008}.
We begin by writing Eq.~(\ref{eq:model}) as a system of first order ODEs:
\begin{equation} \label{gc_compact}
	\dot{\mathbf{x}}=\mathbf{F}\left(t,\mathbf{u},\mathbf{v}\right), \qquad \mathbf{x}= \begin{pmatrix}   \mathbf{u}  \\ \mathbf{v} \end{pmatrix} 
\end{equation}
where  $\mathbf{u}=\left(u_{1},\dots,u_{p}\right)^{T}$ and $\mathbf{v}=\dot{\mathbf{u}}$, respectively,
represent the $N$-dimensional position and velocity vectors. In order to find periodic solutions to this system, we are searching for solutions $\mathbf{x}$ so that $\mathbf{x}(0)=\mathbf{x}(T)$ where $T$ is the fixed period of the solution. 
This suggests that we define the Poincar\'e map: $\mathbf{\cal{P}}(\mathbf{x}^{(0)})=\mathbf{x}^{(0)}-\mathbf{x}(T_{b})$,
where $\mathbf{x}^{(0)}$ is the initial condition and $\mathbf{x}(T_{b})$
is the result of integrating Eq.~(\ref{gc_compact}) forward in time until $t=T_{b}$ using standard ODE integrators \cite{Hairer}. 	
A periodic solution with period $T$ (frequency $1/T$) will be a root of $\mathbf{\cal{P}}$.
To obtain an approximation of this root, one can apply Newton's method \cite[Sec. 3]{Flach2008}.  This method is generally robust in practice, if one is equipped
with a suitable initial guess towards the waveform of
interest, and has been employed
for the computation of breathers, and other time-periodic solutions, in a number
of settings, including Hamiltonian granular chains \cite{Theocharis10} and damped-driven 
ones \cite{hooge12}. An example dark-breather in a damped-driven granular chain found using the numerical procedure just described is shown in Fig.~\ref{breather.2}.  This same
figure also shows the experimental realization of a dark breather in a granular chain.

Another important consideration, especially if experiments are involved, is that of stability. For breathers, the theory is well understood see \cite{Flach2008} and  \cite{Aubry06,Flach05}. Indeed, the linear stability problem is determined in the standard way:
A small perturbation $V(t)$ is added to a solution $\textbf{x}(t)$ that is time-periodic. The result
$  \textbf{x}(t)  + V(t) $ is substituted into Eq.~\eqref{gc_compact} to obtain an
equation describing the evolution of $V$. Keeping 
only linear terms
in $V$ will lead to the so-called variational equations,
\begin{equation} \label{eq:var}
	\dot{V}=\left[\mathrm{D}F\right](t) \,V\,,
\end{equation} 
with initial data $V(0)=\mathbf{I}$; and ${\mathrm{D}F}$ is the Jacobian of the 
equations of motion Eq.~(\ref{gc_compact}) evaluated at the point $\mathbf{x}(t)$
(note that, when computing the Jacobian matrix for Newton's method, one needs to compute
$V(T)$, so in some sense, Newton's method gives stability information ``for free").
Since $\left[{\mathrm{D}F}\right](t)$ will be periodic
in time with period $T$, Eq.~\eqref{eq:var} represents a Hill's equation.  It is well known
within Floquet theory \cite{ODE1}, that the fundamental solutions of Eq.~\eqref{eq:var}
have the property $V(t + T) = \rho V(t) $, where $\rho$ is a so-called Floquet
multiplier (FM). The Floquet multipliers are the eigenvalues of the matrix $V(T)$, where $V(T)$
is the solution of Eq.~\eqref{eq:var} with initial data given by the identity matrix. 
Thus, the perturbation will exhibit exponential growth if there is at least one FM with $|\rho | >1$.  In this case, the corresponding solution $\textbf{x}$ is deemed unstable.
Otherwise, the solution is called spectrally stable. If the lattice is
Hamiltonian, all FMs must lie on the unit circle for the solution to be spectrally 
stable, otherwise the solution is unstable, see \cite[Sec. 4]{Flach2008} for more details. One can connect such 
FMs with more familiar eigenvalue calculations via
the correspondence $\lambda=\log(\rho)/T$, mapping
instability to the real part $\lambda_r>0$.

We have just described a set of powerful tools to study dark breathers
of the granular chain. The tools themselves are not confined, of course,
to dark breathers of the monomer granular chain. Indeed, let us explore
some other relevant examples. If the coefficients $\nu_2,B$ of Eq.~\eqref{NLS}
had the same signs
then the corresponding NLS equation would be focusing,
which has bright soliton solutions (i.e., solutions with a sech form).
In that case, 
Eq.~\eqref{eq:NLSansatz} would yield an approximate bright breather
(here the terms bright simply emphasizes the fact the solution
has tails going to zero). For
any homogeneous chain, one cannot modify the concavity of the dispersion
relation, and hence we must have $\nu_2>0$. The only
way to change the sign of $B$ is to have $0<p<1$, which corresponds
to a so-called strain-softening lattice. Examples of such lattices include
tensegrity~\cite{carpe}, origami~\cite{origami,origami2}, stainless steel cylinders separated 
by polymer foams~\cite{RarefactionNester}, and hollow elliptical cylinders~\cite{HEC_DSW}. 
However, for granular chains, we have
$p=3/2>1$, which leads us to the conclusion that dark breathers are the only breather waveform
possible in the homogeneous granular chain. The nonexistence of bright breathers in the FPUT lattice
(and hence granular chain for small strains) result has been proved rigorously in 
\cite{James2013} (using techniques different than multiple-scale analysis).
Thus, to seek bright discrete breathers in the granular chain, we must
move beyond homogeneous configurations.

\begin{figure}[h]
\centering
    \includegraphics[width=4in]{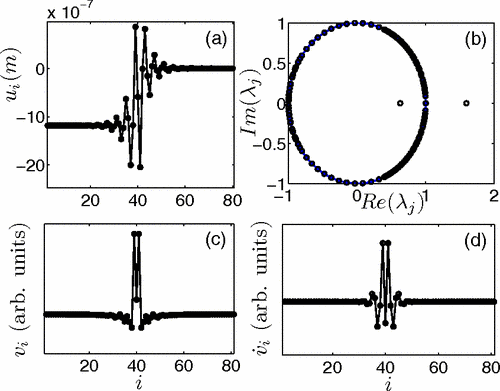}
    \caption{(a) Spatial profile (in displacement)  of a breather of a dimer granular chain.  (b) Corresponding locations of Floquet multipliers in the complex plane. We show the unit circle to guide the eye. Displacement (c) and velocity components (d) of the Floquet eigenvectors associated with the real instability. 
    Figure from Ref. \cite{Theocharis2009}. Used with permission.
    }\label{breather.3}
\end{figure}

If we consider a heterogeneous chain, we can alter the linear dispersion
to our favor. In particular,  dimer chains 
will feature an optical band in the dispersion relation in addition to the acoustic
one found in homogeneous chains. Indeed, if one computes
the dispersion relationship, one finds that the concavity of at the bottom
of the optical band is opposite that of the top of the acoustic band.
According to the previous NLS derivation, this would imply that
$\nu_2 < 0$ (and $B < 0$ since $p>1$) and hence the NLS equation would be
focusing. In this case the profile $Y$ would take the form of a
hyperbolic secant, and hence would be localized in space, i.e.,
it would represent a discrete bright breather. This heuristic argument
can be verified by deriving the NLS equation from the dimer
equations of motion directly, where one indeed finds a focusing NLS equation
when choosing the wavenumber and frequency to be at the bottom of the optical band,
see \cite{Huang2} for a detailed derivation.  The numerical procedure to find
breathers (and compute their stability) was used in \cite{Theocharis10,hooge12} for dimer lattices
and for trimers \cite{hooge11}, see Fig.~\ref{breather.3} for an example. Experiments
for dimers and trimers were reported on in \cite{Boechler2010} and \cite{Stathis}, respectively.  In \cite{pdimer} breathers in a lattice  with
alternating strain-softening and strain-hardening behavior was studied
numerically and analytically. 
These findings clearly indicate how the presence of
heterogeneity once again enables a complete
modification of the monomer GC picture: just like
for TWs, the presence of heterogeneity led to the
complete destruction of perfectly propagating patterns
(aside from isolated mass ratio ``quanta''), similarly
the dimer lattices enable an entirely new
breed of breathers, namely bright breathers.

It is relevant to note here that while GCs have numerous
advantages from an experimental standpoint (including
their distributed visualization, their controllable 
heterogeneity and their controllable ---based on
precompression--- nonlinearity), on-demand initialization
is not one of them. Indeed, while the experimenters
have considerable control over the boundary actuation
of such systems, they do not generally have the ability
to prescribe initial conditions at will. In that sense,
the experimental realization of these breather initial
conditions is not as straightforward of a task. This
was bypassed in~\cite{dark2} through the destructive
interference of two plane waves of suitable frequency
in damped-driven GCs, leading to the spontaneous
emergence of dark breathers. It was also sidetracked
in dimer GCs in the work of~\cite{Boechler2010}
through the use of the nonlinear mechanism of modulational
instability, upon driving the chain at the unstable
frequency of the optical band edge, which spontaneously
was observed to lead to localization in the bandgap
of the linear spectrum.

Breathers cannot only be found in perfectly periodic lattices, like the monomer, dimer, and trimer ones just described, but also in defect lattices too.
From the solid-state physics point of view, this is somewhat natural. When a waveform progating in an otherwise perfect lattice encounters a defect,
that defect particle will be excited and vibrate at the corresponding natural frequency.  In terms of the linear theory, the defect will induce a linear mode
that can possibly lie in the bandgap, e.g., if a
defect mass is introduced that is lighter than
the mass of the rest of the beads~\cite{Defect_Job}. In the presence of nonlinearity, breathing structures can bifurcate from this linear defect mode, and can exist for an interval of frequencies within the gap. There has been a number of studies on defects and corresponding breathing modes in granular chains \cite{Defect_Man,Defect_Job,Theocharis2009}.
These showed also how to naturally interpolate between
the defect limit and the dimer chain discussed above.
Indeed, if there is a single defect mode~\cite{Defect_Job,Theocharis2009}, a single defect mode exists (and a single family of bright breathers accordingly).
If two defect beads are present, say two sites apart, then two defect breathing frequencies (and breather families)
emerge~\cite{Theocharis2009,Defect_Man}. 
As the number of defect beads grows, so does the number of corresponding linear frequencies, until in
the ``infinite defect'' limit, an entire (optical) band
emerges, retrieving the above mentioned limit and its 
bright optical-band-derived breathers.

\section{Dispersive Shock Waves}
\label{DSW}

\begin{figure}[ht]
\begin{minipage}{0.45\textwidth}
    \includegraphics[width=2in]{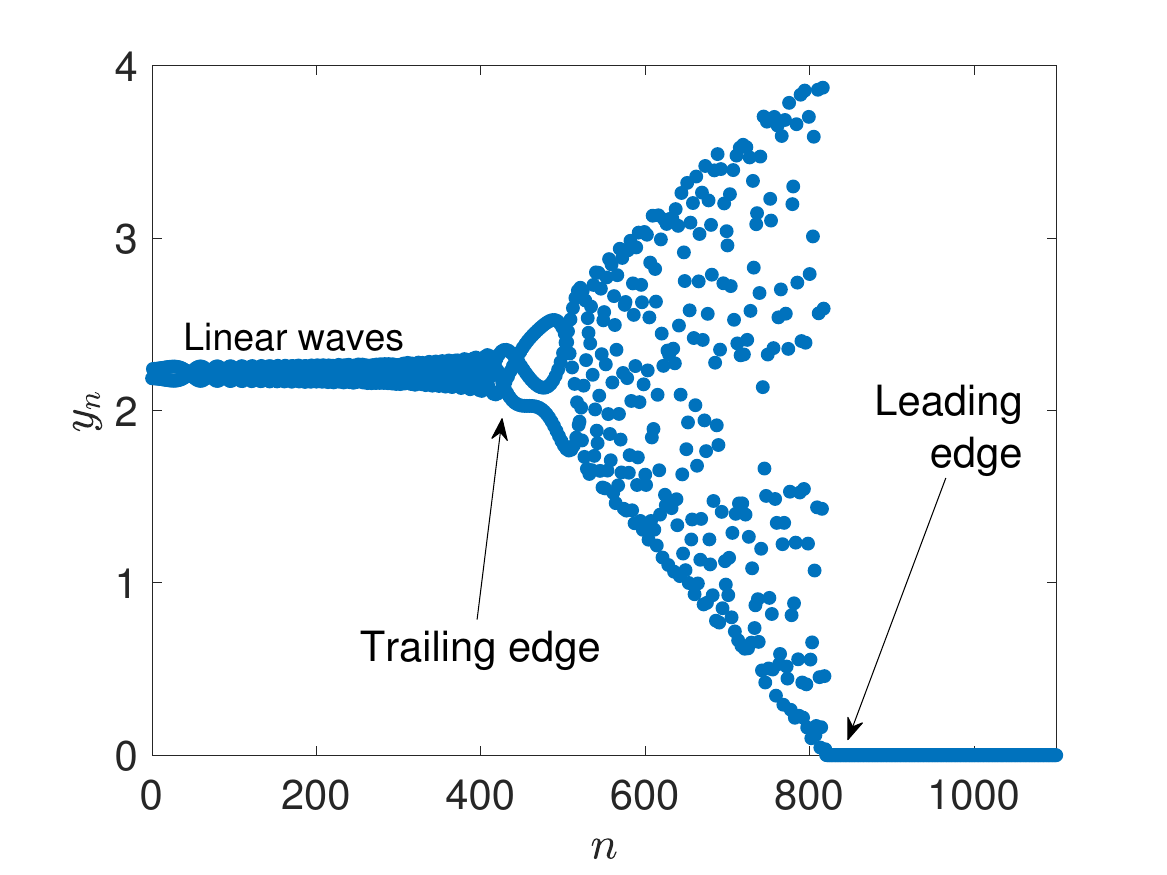}
    \caption{Upon initializing Eq.~\eqref{strain} with step-initial data,
    a DSW forms.}\label{DSW.1}
    \end{minipage}
    \hfill
\begin{minipage}{0.45\textwidth}
    \includegraphics[width=2in]{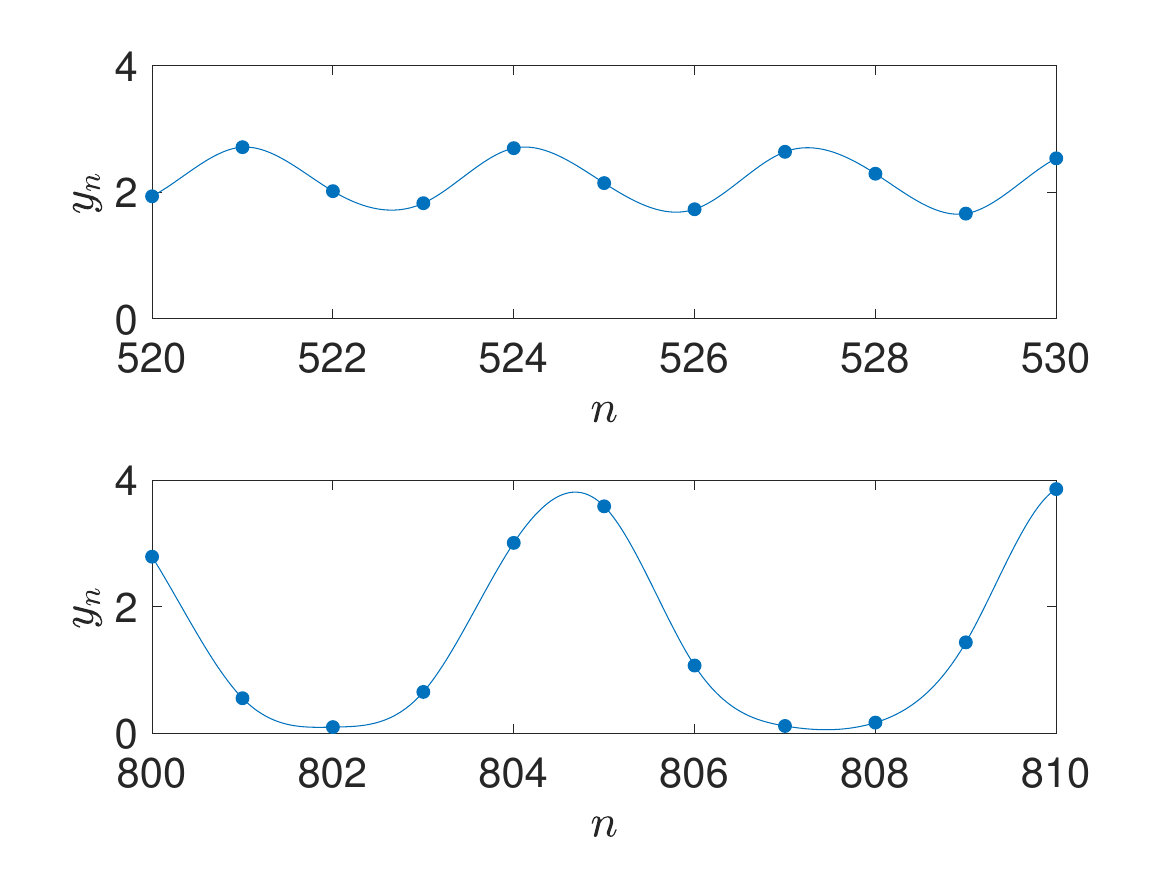}
    \caption{Two different zooms of solution shown in left panel (markers) and interpolating splines (lines). Each pattern appears to be described by a periodic wave, albeit with
    different wavelength, amplitude, and mean.}\label{DSW.2}
\end{minipage}

\end{figure}

What happens to a medium or material when it is subjected to a sudden change of state? Such a question
is important from an engineering perspective, with classic examples including explosions in shock tubes \cite{tube} or when gases or fluids are compressed in a piston chamber \cite{piston}. An idealization of a sudden change of state in the context of nonlinear lattices is represented
in the form of step-initial data, the so-called
Riemann problem,
\begin{equation}\label{step}
y_n(0) = \left\{
\begin{split}
c, \quad & n \leq 0 \\
0, \quad & n > 0
\end{split}
\right. \qquad \dot{y}_n = 0
\end{equation}
For fluid or gas dynamics based on space and time continuous conservation laws, such step initial data leads to the formation of fronts with infinite derivatives
that travel through the medium at constant speed \cite{Smoller}. These are the classical shock waves.
Let us see what happens when we initialize our example nonlinear lattice, Eq.~\eqref{strain},
with step initial data with $c=4$. The result is shown in Fig.~\ref{DSW.1}. Rather than a front
forming, it appears that an oscillating structure connecting states of different amplitude has formed. Evidently, the dispersive nature of Eq.~\eqref{strain} is responsible for the oscillations. For this reason, the structure shown in Fig.~\ref{DSW.1} is a so-called dispersive shock wave (DSW).
Indeed, by inspection, its continuum approximation, see Eq.~\eqref{ch3_tw7},
the lattice equation can be viewed as a conservation law with higher order dispersive corrections
(i.e., the terms pertaining to the higher powers of $\epsilon$). The study of dispersive shock waves in one-dimensional (1D) nonlinear lattices (to be called lattice DSWs) 
dates back several decades ago \cite{first_DSW}, and continues to be
an active area of study from the 1990s~\cite{holian_atomistic_1995} and to
this day~\cite{talcohen,KdVToda_limits}. 
They
have been explored numerically, and even experimentally in several works \cite{Nester2001,Hascoet2000,Herbold07,Molinari2009,shock_trans_granular,HEC_DSW}. 
Although much of the above motivation stems from the granular chains, it is of broader physical interest, as similar
structures have been experimentally observed, e.g., in nonlinear
optics of waveguide arrays~\cite{fleischer2}.
It is important to also highlight in this context 
another setup that has recently emerged, namely tunable magnetic
lattices~\cite{talcohen}. Here, ultraslow shock waves can arise and 
can be experimentally imaged in terms of their space-time
evolution.

To get a sense of how one could analyze a DSW,
let us inspect the solution more closely. Figure.~\ref{DSW.2} shows a zoom of the DSW
near the trailing edge (i.e., the back part of the DSW, see the label in Figure.~\ref{DSW.1})
and near the leading edge (i.e., the front part). At this microscopic level, the solution
appears to be a periodic wave, however the parameters, such as the mean, amplitude, and wavenumber,
seem to depend on ``where" in the lattice one looks. This suggests that the DSW is actually described by a single periodic wave, $\Phi(n -c t)$, whose mean, wavenumber, and amplitude vary slowly across its domain. This is the idea behind modulation theory, pioneered by Whitham over 50 years ago \cite{Whitham74}.
Modulation equations describing the periodic wave parameters can be derived
using, e.g., the method of the so-called averaged Lagrangian~\cite{Whitham74}. 
Equivalent alternative approaches also exist
with, arguably perhaps, the most accessible among
them pertaining to the averaging of the conservation
laws of the model of interest~\cite{Mark2016}.
Such approaches have been successfully employed
to study DSWs in several space-time continuous models, as summarized in \cite{DSWreview,scholar,Mark2016}.
For lattices, however, this approach leads to modulation equations that are quite cumbersome \cite{Venakides99,DHM06}.
A principal reason for this feature can be thought as
the fact that the periodic wave $\Phi$ satisfies an advance-delay equation
(the one we encountered in Eq.~\eqref{step_2}), in which case there is no closed-form solution.
An additional significant complication stems
from the potential absence in the context of
lattices of the feature of a momentum conservation law. While this
is not the case in the FPUT lattices of interest herein,
in general, the lack of integrability (and, more
practically, of a sufficient number of the ``first few''
conservation laws) poses a significant hurdle to
Whitham theory considerations herein.
In such situations, one will often resort to other approximations, by, e.g., focusing
on the trailing or leading edge dynamics \cite{El2005}, employing a harmonic
approximation for the periodic waves \cite{Sprenger2024}, or using a data-driven approach to reduce the dynamics to a planar ODE \cite{CHONG2022}. In this chapter, we will focus
on two integrable limits, in which case tractable analytical approximations
can be obtained, and hence could be useful for practical engineering applications.

We start with the simplest integrable limit, the KdV equation, see Eq.~\eqref{kdv}. Solutions of this KdV equation
that are of critical importance for our purposes are the periodic traveling  waves,
\begin{equation}\label{per}
Y(X,T) = r_1 + r_2 - r_3 + 2(r_3 - r_1) \mathrm{dn}^2 \left(\sqrt{\frac{r_3 - r_1}{6} }(X - V T) ; m \right), 
\end{equation}
where $V = \frac{r_1 + r_2 + r_3}{3}$,  $m = \frac{r_2-r_1}{r_3-r_1}$
and dn is one of the Jacobi elliptic functions with elliptic parameter $0 \leq m \leq 1$ \cite{NIST2010}. Note that these waves are parameterized by $r_1,r_2,r_3$.
Assuming the parameters of the periodic wave (i.e., the $r_j$'s)  vary slowly with respect to $X,T$ and averaging three of the conserved quantities of the KdV equation over a period yields a set of three Whitham modulation equations \cite{Whitham74}. 
In the case of self-similar solutions, $r_j=r_j(X/T)$,
these equations have the form $(S_j - X/T)r'(X/T)=0$, 
where the characteristic speeds $S_j$ are nonlinear functions of $r_1,r_2$ and $r_3$.
Thus, in this self-similar framework, $r_j$ is either constant or $X/T$ is equivalent to the characteristic speed $S_j$. If one assumes the following initial data for the KdV equation
\begin{equation}\label{step2}
Y(X,0 ) = \left\{
\begin{split}
1, \quad & X < 0 \\
0, \quad & X > 0
\end{split}
\right.
\end{equation}
then one finds 
$$r_1 = 0, \qquad r_2 = m, \qquad r_3 = 1. $$
Details for the derivation of these expressions can be found 
in \cite{GP73,Kamchatnov,Mark2016}. 
Substituting these expressions into Eq.~\eqref{per}, and returning to the granular
variables in the strain
 $y_n(t)$,  we obtain the following approximation
for the core of the granular chain DSW:
\begin{equation}\label{eq:granular_kdv_dsw}
y_n(t) =  - c \frac{\sigma^2}{ K_3 }\left(   2\mathrm{dn}^2\left(\sqrt{4 c} \left( n-  \left(1 +   \frac{m+1}{3} c  \right)\sigma t + \Theta_0(n,t) \right); m\right) - (1 - m) \right)
\end{equation}
where $c = \epsilon^2 / 24$ is a small parameter, $\sigma$ is the sound speed and the variables $n,t$ are parameterized 
by $m$ through the expression
\begin{equation}
  \frac{(n-\sigma t)}{\sigma c t} = S(m)\,, 
\end{equation}
with $S(m)$ given by 
\begin{equation} \label{eq:rvel2}
  S(m) = \frac{1+m}{3} - \frac{2}{3} \frac{m(1-m) K(m)}{E(m) - (1-m)K(m)},
\end{equation}
where $K(m)$ and $E(m)$ are complete elliptic integrals of the first and second kind, respectively.
Note Eq.~\eqref{eq:granular_kdv_dsw} contains a slowly modulated phase shift
$\Theta_0(n,t)$ that is not accounted for by the leading order Whitham theory \cite{SIAP59p2162}.
We have included it here, as it will be treated as a fitting parameter to account for a phase
mismatch between theory and simulation. 

From the above expression we can write the trailing edge speed ($n/t=s_-^{\rm KdV}$) and leading edge speed $(n/t=s_+^{\rm KdV})$
in terms of the original granular chain variables,
\begin{subequations}
\label{e:spmkdv}
\begin{eqnarray}
s_-^{\rm KdV} &=&   \sigma \left(1 - c\right) \\
s_+^{\rm KdV} &=&  \sigma\left(1 + \frac{2}{3}c \right).
\end{eqnarray} 
\end{subequations}
Thus the leading edge speed is supersonic, as it necessarily 
lies above the
sound speed $\sigma$, whereas the trailing edge speed is subsonic. Since the parameter
$\epsilon$ is assumed to be small (and hence so is $c$), the leading edge is traveling just
above the sound speed, and the trailing edge is slightly below. 

\begin{figure}[ht]
\begin{minipage}{0.45\textwidth}

    \includegraphics[width=2in]{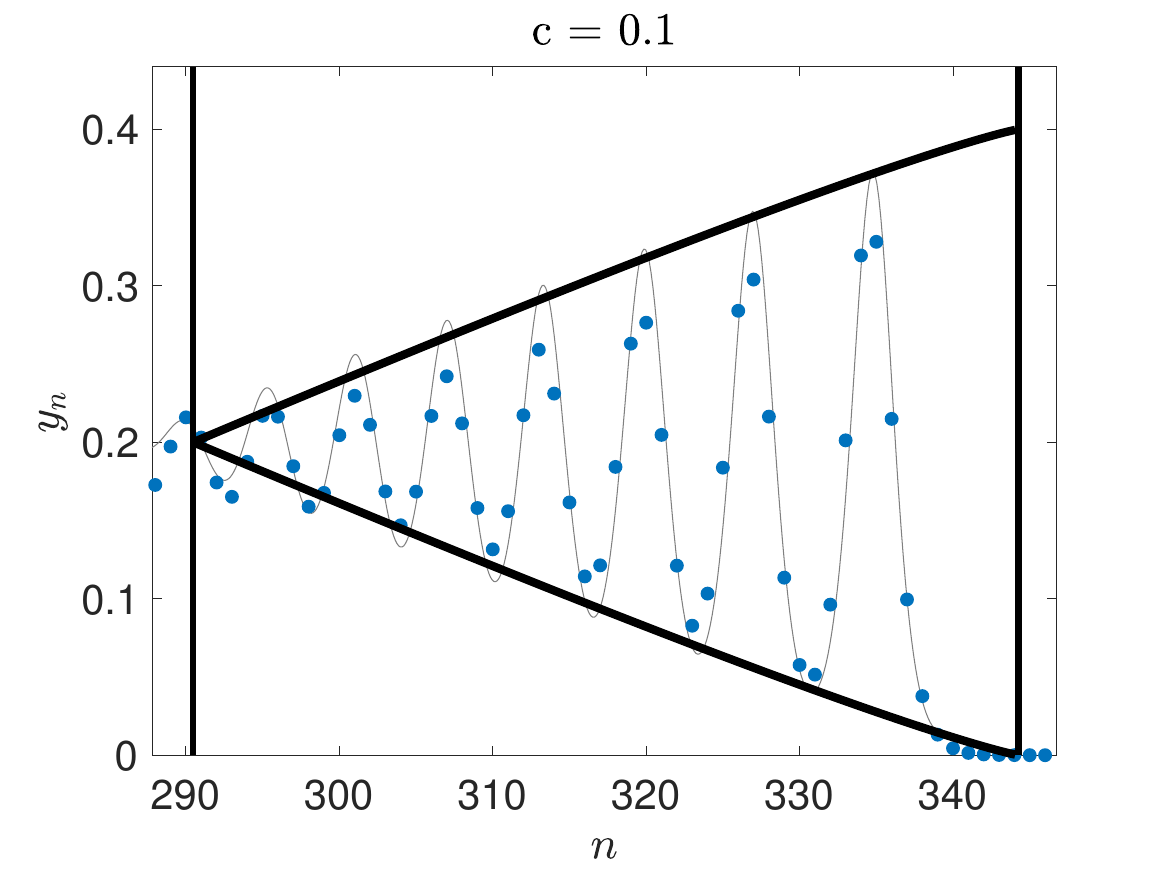}
    \caption{Granular chain simulations with an initial velocity shock
    compared to the KdV prediction with $c = 0.1$.
    Figure from Ref. \cite{CHONG2022}. Used with permission.
    }\label{DSW.3}
    \end{minipage}
    \hfill
\begin{minipage}{0.45\textwidth}
    \includegraphics[width=2in]{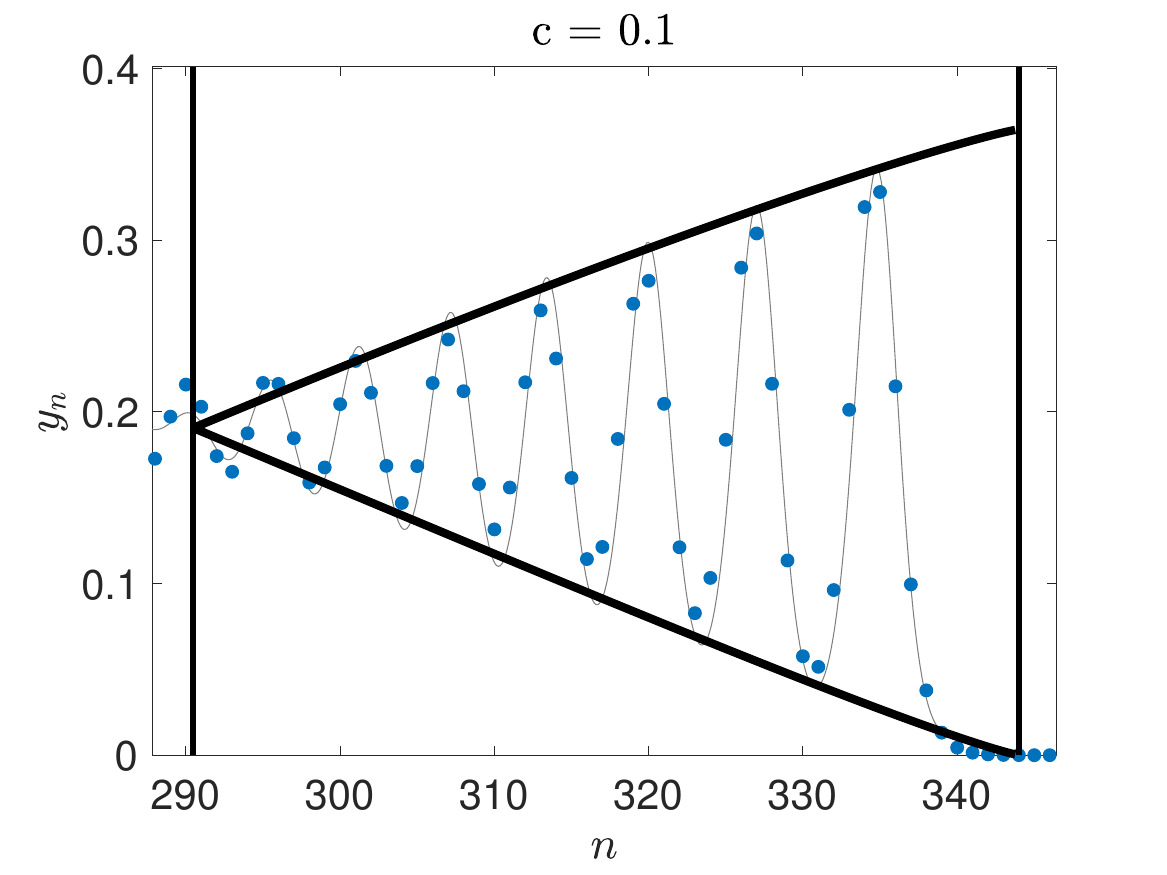}
    \caption{Same as \ref{DSW.3} but the simulation is compared to the Toda
    approximation.Figure from Ref. \cite{CHONG2022}. Used with permission.}\label{DSW.4}
\end{minipage}
\end{figure}

This formulation gives a way to approximate the DSW resulting from
step initial data in the strain, given in Eq.~\eqref{strain}; see \cite{KdVToda_limits}
for comparisons to numerical simulations. However, these initial conditions imply
the entire strain profile must be specified, which is generally hard
to achieve in GC experiments, as discussed above. The more relevant initial condition
for the granular chain, and other similar lattices,  is a collision at
one side of the chain. In \cite{Molinari2009} it was shown that such a collision is
well approximated by a continuous velocity applied to the end of a semi-infinite
chain (this is the so-called piston problem \cite{piston}). By an appropriate change of
variables, the piston-problem initial conditions are equivalent to
an infinite chain with a velocity shock \cite{Venakides99}, namely,
\begin{subequations}
\label{eq:velocity_shock}
\begin{eqnarray}
          u_n(0) &=& 0 \\
      \dot{u}_n(0)& =& - 2 c \, \mathrm{sign}(n)  
\end{eqnarray}  
\end{subequations}
where $\mathrm{sign}(0)=0$. Notice that this initial condition is given in terms of the displacement variable $u_n$. Nonetheless,
we continue to report results in terms of the strain $y_n = u_n - u_{n+1}$ for consistency.
With such an initial condition, it is reasonable to suppose,
based on the linear theory \cite{MielkePatz2017}, that the trailing edge (in the strain) will have a mean close to $2c$. This suggests that
the $c$ defined assuming the initial data Eq.~\eqref{step} is the same
as the $c$ defined in Eq.~\eqref{eq:velocity_shock}. Indeed, this was
the motivation for defining $c=\epsilon^2/24$ previously. This means we can
apply the prediction Eq.~\eqref{eq:granular_kdv_dsw}, even when
the initial condition is given by a velocity shock.  Fig.~\ref{DSW.3} shows a comparison of a DSW
formed given a velocity shock (markers) and the KdV prediction
for  $c=0.1$. 
The microscopic time $t$ is chosen such 
that macroscopic time $T_f = 1200$ is fixed. Figure~\ref{DSW.4} shows the same simulation
data, but now with windowing such that the microscopic variables ($n,t$)
are fixed. In particular, intensity plots of the strain are shown.
The prediction of the leading and trailing speed from the KdV equation
are shown as solid black lines.

\begin{figure}[ht]
\centering
    \includegraphics[width=4in]{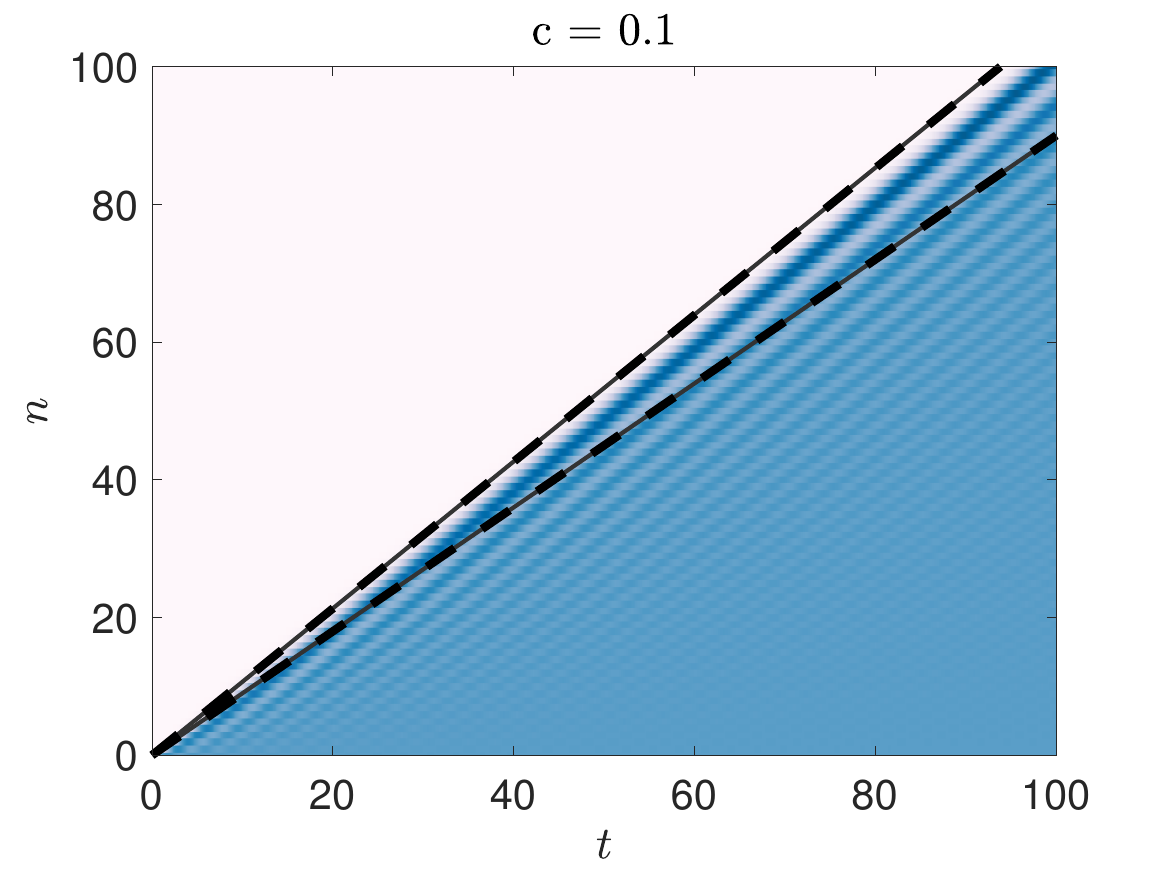}
    \caption{Intensity plot of the granular chain simulation with an initial velocity shock,
    Eq.~\eqref{eq:velocity_shock}. The solid lines are the estimates of the leading
    and trailing edge from the KdV description, and the dashed lines are from the Toda description.
    The small parameter is $c=0.1$.Figure from Ref. \cite{CHONG2022}. Used with permission. }\label{DSW.5}
\end{figure}

We now turn our attention to a different analytical approximation of the granular chain DSW, namely, one based on the Toda lattice, see discussion in Sec.~\ref{TW}.
There is a four parameter family of traveling periodic waves of the Toda lattice,
parameterized by $E_1,E_2,E_3,E_4$.  In terms of the strain, the formula is,
\begin{equation}
y_n(t) = \log\left( 2 \hat R_n(t) + \frac{1}{2}(E_1^2 + E_2^2 + E_3^2 + E_4^2) - \mu_n^2(t) - b_n^2(t)  \right)
\label{e:aelliptic}
\end{equation}
where
\begin{equation} \label{e:belliptic}
    b_n(t) = E_1 + E_2 + E_3 + E_4 - 2\mu_n(t)
\end{equation}
and

\begin{eqnarray*}
    \mu_n(t) &=& E_2 \frac{\displaystyle 1 - (E_1/E_2)\,B\,\mathrm{sn}^2(Z_n(t),m)}{1 - B\,\mathrm{sn}^2(Z_n(t),m)}\,,
\\
Z_n(t)& =& 2n F(\Delta,m) + \omega t\ + Z_0, \label{e:Z}
\\
m &=& \frac{(E_3-E_2)(E_4-E_1)}{(E_4-E_2)(E_3-E_1)}\,,
\label{e:mdef}
\\
\hat R_n(t) &=& -\sigma_n(t) \,\sqrt{P(\mu_n(t))} \\
P(z) &=& (z-E_1)(z-E_2)(z-E_3)(z-E_4)\,,
\\
\omega& =& \sqrt{(E_4-E_2)(E_3-E_1)} \\
\Delta &=& \sqrt{\frac{E_4-E_2}{E_4-E_1}}\,,\quad
B = \frac{E_3-E_2}{E_3-E_1}\,.
\end{eqnarray*}

In the above equations,
$\mathrm{sn}(z,m)$ is the Jacobi elliptic sine, 
$F(z,m)$ is the inverse of $\mathrm{sn}(z,m)$, 
$Z_0$ is an arbitrary translation parameter (phase), $\mu_n(t)$ is the Dirichlet eigenvalue of the scattering problem for the Toda lattice
and $\sigma_n(t) = \pm1$ is the sign associated with $\mu_n(t)$, and determines whether $\mu_n(t)$ is increasing or decreasing as a function of $n$ and $t$. For numerical computations,
we used $\sigma_n(t) = -\mathrm{sign}( \mod(Z_n(t)/K_m) - 1/2   )$.
Notice how the Toda traveling wave solution is more complicated
than its KdV counterpart in Eq.~\eqref{per}. So while we may anticipate a better approximation
(since no long-wavelength assumption is made), the 
tradeoff is a formula that will be considerably more cumbersome.

Like in the KdV case,
one can derive a system of modulation equations (4 in the Toda case) that can be written in diagonalized form, and solved in the self-similar frame $E_j=E_j(n/t)$ assuming an initial velocity shock. Three of the parameters
are constant in the self-similar frame, and one depends on the parameter
$m$ \cite{blochkodama,Gino2023}. In particular,
\begin{equation} \label{eq:Es}
    E_1  = -(1+c), \quad
E_2  = -(c-1), \quad 
E_3(m) =   (c+1) \frac{1-c(1-m)}{1+c(1-m)}, \quad
E_4 = (c+1). 
\end{equation}
Assuming $0<c<1$, the core of the DSW
is described by Eq.~\eqref{e:aelliptic} with parameters given via
Eq.~\eqref{eq:Es}
where $n,t$ are parameterized by $m$ through the expression
\begin{equation} \label{eq:Toda_vel}
\frac{c(c+1)}{1+c(1-m)}
  \frac{(1+c(1-m)) E(m) - (c+1)(1-m) K(m)}{(c+1) K(m) - (1+c(1-m))\Pi \left(\left.\frac{cm}{c+1}\right|m\right)} =: s^{\rm Toda}(m)  = \frac{n}{t}  
\end{equation}
where $K(m)$, $E(m)$ and $\Pi(n|m)$ are complete elliptic integrals of the first, second, and third kind respectively, and $s^{\rm Toda}(m)$
is the third charecteristic velocity of the Whitham equations for the
Toda system. As before, $m\rightarrow 0$ corresponds to the trailing, harmonic wave edge (recall $0<c<1$) and 
$m\rightarrow 1$ corresponding to the leading, solitary wave edge.
The trailing edge speed ($s_-^{\rm Toda}$) and leading edge speed ($s_+^{\rm Toda}$)  are obtained as limiting values in Eq.~\eqref{eq:Toda_vel}, in particular
\begin{equation} \label{eq:Toda_edgespeeds}
 \lim_{m\rightarrow 0} s^{\rm Toda}(m) =s_-^{\rm Toda} = 1-c, \qquad \lim_{m\rightarrow 1} s^{\rm Toda}(m) = s_+^{\rm Toda} =  \frac{\sqrt{c(c+1)}}{\log(\sqrt{c}+\sqrt{c+1})}
\end{equation}
Once again, we have that the leading edge speed is supersonic whereas the trailing edge speed is subsonic. 
We can also compute the trailing edge mean and leading edge amplitude from Eq.~\eqref{e:aelliptic},
\begin{equation} \label{eq:Toda_meanamp}
\bar{y}_-^{\rm Toda} = 2\log(1+c), \qquad a_+^{\rm Toda} = 2 \log(1 + 2c)
\end{equation}
Notice that trailing edge speed predictions from Toda and KdV are identical, namely $s_-^{\rm Toda} = s_-^{\rm KdV}$.
Indeed, the remaining three edge characteristics are also related. By
Taylor expanding the remaining formulas in Eqs.~\eqref{eq:Toda_edgespeeds} and \eqref{eq:Toda_meanamp} about $c=0$ shows that the leading order behavior is identical to the KdV approximation. Namely, for small $c$
we have that
\begin{equation} \label{eq:compare_parms}
s_+^{\rm Toda} \approx 1 + \frac{2}{3} c = s_+^{\rm KdV}, \qquad \bar{y}_-^{\rm Toda} \approx 2c = \bar{y}_-^{\rm KdV},\qquad
a_+^{\rm Toda} \approx 4c = a_+^{\rm KdV} 
\end{equation}
Thus, to leading order, both the KdV and Toda limits yield the same approximation
of the edge characteristics. 

It is important once again to put these results
in the proper context in the setting of GCs. What we
have done herein is that we have leveraged a mapping
through multiple scales (in the KdV case) or a matching
(up to the maximal allowable order) in the Toda case
of the GC problem {\it with precompression} to
well-established integrable analogs for which
the rich structure of integrability (the periodic
solutions, the conservation laws, the Lagrangian/Hamiltonian
structure, the Riemann invariants, etc.) enable
the completion of the Whitham equation ``program'' and
therefore the approximate, yet quite satisfactory description of the GC DSWs. On the other hand, we
have not addressed the much more complex case of the
so-called sonic vacuum, i.e., when the precompression
$\delta_0=0$. Yet, it was in the latter limit that
the principal earlier studies~\cite{Herbold07,shock_trans_granular,Molinari2009}
were conducted. Naturally, this raises the question
of how to address the latter limit. It turns
out that some of the asymptotic techniques for
describing the trailing and leading edges of DSWs,
such as~\cite{El2005}, can be carried out in 
discrete as well as continuous systems and can be
brought to bear in the GC without precompression.
Yet, in order to perform a systematic analysis
of both limits and to obtain a framework for
Whitham modulation theory a quasi-continuum
methodology can be proposed. While those of~\cite{Nester2001} and~\cite{pikovsky} suffer
 from issues of well-posedness, regularized variants
 thereof in the spirit of Rosenau's work~\cite{rosenau1,rosenau2} bypass this concern and
 are promising grounds for developing a sonic-vacuum DSW description. This is an interesting direction of ongoing work~\cite{suyang}.

\section{Summary and Future Directions}
\label{SFD}
In this chapter we summarized some of the key tools to study nonlinear lattices. We used the granular chain as the 
 vehicle of choice, formulating the mathematical
equation framework (of FPUT systems) in order to demonstrate how these tools, such as continuum modeling, Fourier analysis, and various numerical methods could be applied. Along this short journey, we sought to highlight some of the main results concerning granular chains revolving
around three fundamental structures, namely the solitary waves, the breathers, and the dispersive shock waves. In this section, we touch upon some emerging directions in the realm of nonlinear lattices.

While the granular chain serves as a canonical example of a nonlinear lattice,
there exists a variety of other platforms that can be modeled
in the general framework of~\eqref{eq:model} upon modifying existing, or adding additional, terms.
We have already discussed periodic arrangements and defect lattices,
but there exist also systems incorporating disorder, where effects such as Anderson localization
are possible \cite{Martinez_disorder_2016},
and superdiffusive transport has been realized~\cite{Kim_disorder_2016}. Additionally, systems with an onsite force have also been explored.
Such examples include the study of breathers in the Newton's cradle \cite{JAMES201339}
and also the exploration of~\cite{Liu2015,Liu2016}.
Traveling waves as well
as nanoptera (waves with small amplitude tails) were predicted in dimers~\cite{faver_dimer,perline}, as well as in locally resonant systems~\cite{yuli,Stefanov2015} (in the latter, in the setting of the so-called woodpile 
chains, they were also observed~\cite{jkprl}); see also the recent review~\cite{VAINCHTEIN2022133252}. One can also
consider the intersite force function
to be of a slightly more general form, namely $F(x) = A(d+ \mathrm{sign(p)} x)^p$.
The case of $0<p<1$ corresponds to a so-called strain-softening lattice. 
In such lattices, traveling waves have been studied in the context of
tensegrity~\cite{carpe}, origami chains~\cite{origami,origami2} and
stainless steel cylinders separated 
by polymer foams~\cite{RarefactionNester}, whereas dispersive
shock waves have been explored in hollow elliptical cylinders~\cite{HEC_DSW}.
Breathers in lattices that alternate between softening and hardening were explored recently
in \cite{pdimer}. An array of repelling magnets corresponds to the case with
$p<0$, in which case traveling waves \cite{moleron}, and breathers
in one \cite{magBreathers} and two dimensions \cite{magBreathers2D} have been studied, as well
as DSWs \cite{talcohen}. Magnetic lattices with time-modulated stiffness properties, falling into the class of so-called phononic time crystals,
have also been studied, where e.g., time-periodic
and wavenumber bandgap breathers have been explored \cite{Kim23,Kim24}.

Magnetic lattices constitute a particularly interesting
example, as they also include long-range interactions,
whereby each neighbor is coupled to all other neighbors, albeit weakly.
In \cite{magBreathers}, breathers with algebraically decaying tails in
lattices with long-range interactions were shown to exist, confirming
theoretical predictions from decades prior in \cite{FlachLong}.
The notion of long-range media is intimately connected to
that of fractional media. Fractional media, where the underlying 
model equations have the usual (integer) derivatives replaced
with fractional ones (based on suitable integral definitions), are relevant in many areas,
such as fluid and quantum mechanics, and more \cite{FractionalBook}. In the latter book, connections
have been made to fractional discrete variants
of the nonlinear Schr{\"o}dinger and sine-Gordon
equations, suggesting that a near-future analogue
will be FPUT lattices, that, in turn, will be 
connected with fractional KdV continuum models (also
studied therein).
Fractional media in nonlinear lattices, such as electrical ones
and those of the DNLS type have also been recently proposed \cite{Molina21},
where the the fractional derivatives tend to, among other effects, be tantamount to long-range interactions which can
accordingly, e.g., decrease the threshold for the self-trapping transitions, induce bistability, etc.
We expect this to be a direction that is further explored
in future studies; it also remains to be seen whether,
in addition to spatial fractional derivatives, temporal
fractional derivatives such as the Caputo derivative
for wave-like systems may be of practical interest
(in addition to being of mathematical one)~\cite{maciasbountis2022}.

Band topology, an inherently linear phenomenon, has been explored in electronic materials, optical lattices in cold-atom systems, and topological photonics 
where the behavior in the presence of corners, edges, and surfaces 
has been explored \cite{topphoto,topological}.  
The ability of the relevant topological systems to
preserve features under continuous deformation
presents exceptional promise for guiding of
wave propagation~\cite{ni2023topological}, while
demonstrating robustness against various imperfections.
It is for that reason that relevant applications
have recently extended to both acoustics~\cite{zhang2018topological,xue2022topological}, and mechanics~\cite{xin2020topological,xue2022topological}.
While important steps have been made as concerns the interplay of nonlinearity and novel linear states (such as the topological ones just mentioned) in lattices, e.g.,
in the context of optical waveguides, as summarized in the recent review of~\cite{ABLOWITZ2022133440}, we expect
this direction to be one warranting significant 
additional progress in the near future.

Another exotic linear phenomenon concerns the presence
of
the so-called flat bands, which arise as completely degenerate energy bands in certain tight-binding lattices with macroscopic degeneracy~\cite{leykam2013flat, leykam2018artificial}. 
This phenomenon has  significant implications
such as the associated vanishing group velocity,
which, in turn, suppresses transmission, while
the system produces (generically, as it can be proved)
compactly supported localized (linear) states.
These can be indeed localized with a strictly finite
support. The particular sensitivity of such bands
to perturbation effects enables a wide range
of intriguing features including localization-delocalization
transitions, symmetry-breakings, compact nonlinear
breathing or localized states, ferromagnetic and
superfluid-like features among others.
Once again, the versatility of dispersive platforms
has enabled the realization of such states
in photonic, polaritonic, electrical, magnonic
and even phononic 2D~\cite{phononic2d}
and acoustic 3D~\cite{acoustic3d} systems.
Once again, the interplay of this key linear feature
with nonlinearity is far less explored; one 
recent experimental example is in the electrical lattice
realm of~\cite{FlatbandElectric}, where nonlinear
compactly supported breathing states for diamond
and stub lattices were realized. Further exploration
of the fascinating features of Kagom{\'e}~\cite{kagomejoh},
Lieb lattices~\cite{mukherjee2015observation,vicencio2015observation} and other similar settings~\cite{morales2016simple}
is certainly worthwhile as well.

Most of the considerations listed above are in the realm of one-dimensional lattices,
and indeed the vast majority of studies have been in this setting. The
effect of dimensionality can be significant, however. 
For example, in square or hexagonal lattices, a single particle is in contact with multiple particles. Thus, if energy is applied to a single particle,
it will be distributed to an expanding array of neighbors \cite{andrea}. 
This spreading of energy should prevent the possibility of genuine 2D traveling solitary waves in such lattices.
There have been, however, several studies related to the notion of traveling waves. These include those focusing on the dynamics after impacts, networks of 1D chains that live in 3D space, and ``sound bullets'' which are waves that can be guided to (focus on) preselected lattice sites within the crystal. Such findings are summarized in \cite{granularBook}.
While the geometry of the higher dimensional lattices seems to prohibit genuine traveling solitary waves, it still affords
the possibility of breathers. Indeed, they have been studied
in hexagonally arranged magnetic and
other related lattices, see \cite{magBreathers2D} and references therein. Dispersive shock waves
in 2D lattices should also be possible, however they remain rather unexplored.  Furthermore, it is in such
higher dimensional settings where the effect
and robustness of topological modes could
prove especially versatile; while 
important steps have been made in this direction~\cite{ruzze}, numerous further insights
promise to arise along this vein.

Finally, modern computational approaches involving data driven methods have just started to make an impact on field of nonlinear waves. For example,
recently in \cite{DMD_topological,DMD_topological2} topological states in elastic media were explored
through the lens of dynamic mode decomposition,  which is a dimensionality reduction method to create reduced-order models which identifies spatiotemporal coherent structures from high-dimensional data \cite{colbrook2023multiverse}.
While such methods have been applied to a host of other scientific fields
such as fluid dynamics, control, robotics, and biological science \cite{DMD_topological} their application in the realm of nonlinear lattices
remains in its infancy and thus serves as a crucial future direction in the field. Moreover, the identification
and modeling of features that have long
remained controversial (such as the inclusion of
dissipation; see the discussion of~\cite{granularBook,James_2021}) could benefit from
the emerging boom of data-driven physics-informed,
sparse identification techniques
such as PINNs~\cite{raissi_physics-informed_2019}, SINDy~\cite{brunton_discovering_2016}, deepXDE~\cite{lu2021deepxde} (see also the
review of~\cite{karniadakis2021physics}). 
Such techniques are only starting to be used
to discover mathematical and potentially physically-interpretable-features from data in nonlinear 
dynamical lattices
(see, e.g.,~\cite{SAQLAIN2023107498} for a recent example)
and progress in this direction will, undoubtedly
(in our view), significantly impact  future 
discoveries in nonlinear lattice dynamics.

\bibliographystyle{unsrt}

\bibliography{Chong}

\begin{thebibliography}{100}

\bibitem{LEDERER20081}
Falk Lederer, George~I. Stegeman, Demetri~N. Christodoulides, Gaetano Assanto,
  Moti Segev, and Yaron Silberberg.
\newblock Discrete solitons in optics.
\newblock {\em Physics Reports}, 463(1):1--126, 2008.

\bibitem{RevModPhys.78.179}
Oliver Morsch and Markus Oberthaler.
\newblock Dynamics of bose-einstein condensates in optical lattices.
\newblock {\em Rev. Mod. Phys.}, 78:179--215, Feb 2006.

\bibitem{Peybi}
M.~Peyrard.
\newblock Nonlinear dynamics and statistical physics of {DNA}.
\newblock {\em Nonlinearity}, 17:R1, 2004.

\bibitem{remoissenet}
M.~Remoissenet.
\newblock {\em Waves Called Solitons}.
\newblock Springer-Verlag, Berlin, 1999.

\bibitem{willis}
S.~Flach and C.R. Willis.
\newblock Discrete breathers.
\newblock {\em Physics Reports}, 295(5):181--264, 1998.

\bibitem{Flach2008}
Sergej Flach and Andrey~V. Gorbach.
\newblock {Discrete breathers — Advances in theory and applications}.
\newblock {\em Physics Reports}, 467(1-3):1--116, oct 2008.

\bibitem{Aubry06}
S.~Aubry.
\newblock Discrete breathers: {L}ocalization and transfer of energy in discrete
  {H}amiltonian nonlinear systems.
\newblock {\em Physica D}, 216:1, 2006.

\bibitem{FPUreview}
G.~Gallavotti.
\newblock {\em The Fermi--Pasta--Ulam Problem: A Status Report}.
\newblock Springer-Verlag, Berlin, Germany, 2008.

\bibitem{kev09}
P.G. Kevrekidis.
\newblock {\em {The discrete nonlinear Schr{\"o}dinger Equation}}.
\newblock Springer-Verlag, Heidelberg, 1st edition, 2009.

\bibitem{DPbook}
D.~Pelinovsky.
\newblock {\em Localization in Periodic Potentials: From Schr\"odinger
  Operators to the Gross--Pitaevskii Equation}.
\newblock Cambridge University Press, Cambridge, 2011.

\bibitem{floyd}
J.~Cuevas, P.G. Kevrekidis, and F.L. Williams.
\newblock {\em {The sine-Gordon Model and its Applications: From Pendula and
  Josephson Junctions to Gravity and High Energy Physics}}.
\newblock Springer-Verlag, Heidelberg, 2014.

\bibitem{p4book}
J.~Cuevas-Maraver and P.G. Kevrekidis~(eds.).
\newblock {\em A dynamical perspective on the $\phi^4$ model}.
\newblock Nonlinear Systems and Complexity. Springer International Publishing,
  1st edition, 2019.

\bibitem{granularBook}
C.~Chong and P.~G. Kevrekidis.
\newblock {\em Coherent Structures in Granular Crystals: From Experiment and
  Modelling to Computation and Mathematical Analysis}.
\newblock Springer, New York, 2018.

\bibitem{Mark2016}
G.A. El and M.A. Hoefer.
\newblock Dispersive shock waves and modulation theory.
\newblock {\em Physica D: Nonlinear Phenomena}, 333:11, 2016.

\bibitem{Nester2001}
V.F. Nesterenko.
\newblock {\em Dynamics of Heterogeneous Materials}.
\newblock Springer-Verlag, New York, 2001.

\bibitem{sen08}
S.~Sen, J.~Hong, J.~Bang, E.~Avalos, and R.~Doney.
\newblock Solitary waves in the granular chain.
\newblock {\em Phys. Rep.}, 462:21, 2008.

\bibitem{vakakis_review}
A.~F. Vakakis.
\newblock Analytical methodologies for nonlinear periodic media.
\newblock In {\em Wave Propagation in Linear and Nonlinear Periodic Media
  (International Center for Mechanical Sciences (CISM) Courses and Lectures)},
  page 257. Springer-Verlag, Berlin, Germany, 2012.

\bibitem{yuli_book}
Yu. Starosvetsky, K.R. Jayaprakash, M.~Arif Hasan, and A.F. Vakakis.
\newblock {\em Dynamics and Acoustics of Ordered Granular Media}.
\newblock World Scientific, Singapore, 2017.

\bibitem{Nester_dimer}
S.~Y. Wang and V.~F. Nesterenko.
\newblock Attenuation of short strongly nonlinear stress pulses in dissipative
  granular chains.
\newblock {\em Phys. Rev. E}, 91:062211, 2015.

\bibitem{Spadoni}
A.~Spadoni and C.~Daraio.
\newblock Generation and control of sound bullets with a nonlinear acoustic
  lens.
\newblock {\em Proc. Natl. Acad. Sci. USA}, 107:7230, 2010.

\bibitem{Nature11}
N.~Boechler, G.~Theocharis, and C.~Daraio.
\newblock Bifurcation-based acoustic switching and rectification.
\newblock {\em Nat. Mater.}, 10(9):665--668, 2011.

\bibitem{theocharis_review}
G.~Theocharis, N.~Boechler, and C.~Daraio.
\newblock Nonlinear phononic periodic structures and granular crystals.
\newblock In {\em Acoustic Metamaterials and Phononic Crystals}, pages
  217--251. Springer-Verlag, Berlin, Germany, 2013.

\bibitem{Li_switch}
F.~Li, P.~Anzel, J.~Yang, P.G. Kevrekidis, and C.~Daraio.
\newblock Granular acoustic switches and logic elements.
\newblock {\em Nat. Comm.}, 5:5311--5315, 2014.

\bibitem{bone}
Jinkyu Yang, Sophia~N. Sangiorgio, Sean~L. Borkowski, Claudio Silvestro, Luigi
  De~Nardo, Chiara Daraio, and Edward Ebramzadeh.
\newblock {Site-Specific Quantification of Bone Quality Using Highly Nonlinear
  Solitary Waves}.
\newblock {\em Journal of Biomechanical Engineering}, 134(10), 10 2012.
\newblock 101001.

\bibitem{Hertz}
H.~Hertz.
\newblock {\"Uber die Ber\"uhrung fester elastischer K\"orper}.
\newblock {\em J. Reine. Angew. Math}, 92:156, 1881.

\bibitem{Johnson}
K.~L. Johnson.
\newblock {\em Contact Mechanics}.
\newblock Cambridge University Press, Cambridge, UK, 1985.

\bibitem{FPU55}
E.~Fermi, J.~Pasta, and S.~Ulam.
\newblock {Studies of Nonlinear Problems. I.}
\newblock {\em Tech. Rep.}, (Los Alamos National Laboratory, Los Alamos, NM,
  USA):LA--1940, 1955.

\bibitem{zabu}
Mason~A. Porter, Norman~J. Zabusky, Bambi Hu, and David~K. Campbell.
\newblock Fermi, pasta, ulam and the birth of experimental mathematics: A
  numerical experiment that enrico fermi, john pasta, and stanislaw ulam
  reported 54 years ago continues to inspire discovery.
\newblock {\em American Scientist}, 97(3):214--221, 2009.

\bibitem{daux}
Thierry Dauxois.
\newblock {Fermi, Pasta, Ulam, and a mysterious lady}.
\newblock {\em Physics Today}, 61(1):55--57, 01 2008.

\bibitem{pegof2}
G.~Friesecke and R.~L. Pego.
\newblock Solitary waves on {Fermi}--{Pasta}--{Ulam} lattices: {I}. {L}inear
  implies nonlinear stability.
\newblock {\em Nonlinearity}, 15:1343, 2002.

\bibitem{pegof3}
G.~Friesecke and R.~L. Pego.
\newblock Solitary waves on {Fermi}--{Pasta}--{Ulam} lattices: {III}.
  {H}owland-type {F}loquet theory.
\newblock {\em Nonlinearity}, 17:207, 2004.

\bibitem{pegof4}
G.~Friesecke and R.~L. Pego.
\newblock Solitary waves on {Fermi}--{Pasta}--{Ulam} lattices: {IV}. {P}roof of
  stability at low energy.
\newblock {\em Nonlinearity}, 17:229, 2004.

\bibitem{Shen}
Y.~Shen, P.~G. Kevrekidis, S.~Sen, and A.~Hoffman.
\newblock Characterizing traveling-wave collisions in granular chains starting
  from integrable limits: {T}he case of the {K}orteweg--de {V}ries equation and
  the {T}oda lattice.
\newblock {\em Phys. Rev. E}, 90:022905, 2014.

\bibitem{Rosenau93}
Philip Rosenau and James~M. Hyman.
\newblock Compactons: Solitons with finite wavelength.
\newblock {\em Phys. Rev. Lett.}, 70:564--567, 1993.

\bibitem{Coste}
C.~Coste, E.~Falcon, and S.~Fauve.
\newblock Solitary waves in a chain of beads under {H}ertz contact.
\newblock {\em Phys. Rev. E}, 56:6104, 1997.

\bibitem{pikovsky}
Karsten Ahnert and Arkady Pikovsky.
\newblock Compactons and chaos in strongly nonlinear lattices.
\newblock {\em Phys. Rev. E}, 79:026209, 2009.

\bibitem{MacKay99}
R.~S. MacKay.
\newblock Solitary waves in a chain of beads under {H}ertz contact.
\newblock {\em Phys. Lett. A}, 251:191, 1999.

\bibitem{Wattis}
G.~Friesecke and J.~A.~D. Wattis.
\newblock Existence theorem for solitary waves on lattices.
\newblock {\em Commun. Math. Phys.}, 161:391, 1994.

\bibitem{JKdimer}
E.~Kim, R.~Chaunsali, H.~Xu, J.~Castillo, J.~Yang, P.~G. Kevrekidis, and A.~F.
  Vakakis.
\newblock Nonlinear low-to-high frequency energy cascades in diatomic granular
  crystals.
\newblock {\em Phys. Rev. E}, 92:062201, 2015.

\bibitem{Chatterjee}
A.~Chatterjee.
\newblock Asymptotic solution for solitary waves in a chain of elastic spheres.
\newblock {\em Phys. Rev. E}, 59:5912, 1999.

\bibitem{pego1}
Jared~M. English and Robert~L. Pego.
\newblock On the solitary wave pulse in a chain of beads.
\newblock {\em Proceedings of the AMS}, 133:1763, 2005.

\bibitem{Hochstrasse}
D.~Hochstrasser, F.G. Mertens, and H.~{B\"uttner}.
\newblock An iterative method for the calculation of narrow solitary
  excitations on atomic chains.
\newblock {\em Physica D}, 35:259, 1989.

\bibitem{Jayaprakash1}
K.~R. Jayaprakash, Yuli Starosvetsky, and Alexander~F. Vakakis.
\newblock New family of solitary waves in granular dimer chains with no
  precompression.
\newblock {\em Phys. Rev. E}, 83:036606, 2011.

\bibitem{mason1}
M.~A. Porter, C.~Daraio, E.~B. Herbold, I.~Szelengowicz, and P.~G. Kevrekidis.
\newblock Highly nonlinear solitary waves in periodic dimer granular chains.
\newblock {\em Phys. Rev. E}, 77:015601(R), 2008.

\bibitem{mason2}
M.~A. Porter, C.~Daraio, I.~Szelengowicz, E.~B. Herbold, and P.~G. Kevrekidis.
\newblock Highly nonlinear solitary waves in heterogeneous periodic granular
  media.
\newblock {\em Physica D}, 238:666, 2009.

\bibitem{Jayaprakash2}
K.~R. Jayaprakash, Y.~Starosvetsky, A.~F. Vakakis, and O.~V. Gendelman.
\newblock Nonlinear resonances leading to strong pulse attenuation in granular
  dimer chains.
\newblock {\em J. Nonlinear Sci.}, 23(3):363, 2013.

\bibitem{vak_star3}
K.R. Jayaprakash, A.~Shiffer, and Y.~Starosvetsky.
\newblock Traveling waves in trimer granular lattice {I}: Bifurcation structure
  of traveling waves in the unit-cell model.
\newblock {\em Communications in Nonlinear Science and Numerical Simulation},
  38:8 -- 22, 2016.

\bibitem{staros2}
K.~R. Jayaprakash, Alexander~F. Vakakis, and Yuli Starosvetsky.
\newblock Solitary waves in a general class of granular dimer chains.
\newblock {\em J. App. Phys.}, 112(3):034903, 2012.

\bibitem{perline}
Anna Vainchtein, Yuli Starosvetsky, J.~Douglas Wright, and Ron Perline.
\newblock Solitary waves in diatomic chains.
\newblock {\em Phys. Rev. E}, 93:042210, Apr 2016.

\bibitem{faver}
T.~E. Faver and J.~D. Wright.
\newblock Exact diatomic {F}ermi--{P}asta--{U}lam--{T}singou solitary waves
  with optical band ripples at infinity.
\newblock {\em arXiv:1511.00942}, 2015.

\bibitem{morris}
R.K. Dodd, J.C. Eilbeck, J.D. Gibbon, and H.C. Morris.
\newblock {\em Solitons and Nonlinear Wave Equations}.
\newblock Academic Press, San Diego, 1982.

\bibitem{ST}
A.~J. Sievers and S.~Takeno.
\newblock Intrinsic localized modes in anharmonic crystals.
\newblock {\em Phys. Rev. Lett.}, 61:970--973, Aug 1988.

\bibitem{page}
J.~B. Page.
\newblock Asymptotic solutions for localized vibrational modes in strongly
  anharmonic periodic systems.
\newblock {\em Phys. Rev. B}, 41:7835--7838, 1990.

\bibitem{macaub}
R~S MacKay and S~Aubry.
\newblock Proof of existence of breathers for time-reversible or {Hamiltonian}
  networks of weakly coupled oscillators.
\newblock {\em Nonlinearity}, 7(6):1623, 1994.

\bibitem{dark}
C.~Chong, P.~G. Kevrekidis, G.~Theocharis, and Chiara Daraio.
\newblock Dark breathers in granular crystals.
\newblock {\em Phys. Rev. E}, 87:042202, 2013.

\bibitem{dark2}
C.~Chong, F.~Li, J.~Yang, M.~O. Williams, I.~G. Kevrekidis, P.~G. Kevrekidis,
  and C.~Daraio.
\newblock Damped-driven granular chains: {A}n ideal playground for dark
  breathers and multibreathers.
\newblock {\em Phys. Rev. E}, 89:032924, 2014.

\bibitem{huang1}
Guoxiang Huang, Zhu-Pei Shi, and Zaixin Xu.
\newblock Asymmetric intrinsic localized modes in a homogeneous lattice with
  cubic and quartic anharmonicity.
\newblock {\em Phys. Rev. B}, 47:14561, 1993.

\bibitem{DarkBook}
P.~G. Kevrekidis, D.~J. Frantzeskakis, and R.~Carretero-Gonz{\'a}lez.
\newblock {\em The Defocusing Nonlinear Schr\"odinger Equation}.
\newblock SIAM, Philadelphia, 2015.

\bibitem{Hairer}
E.~Hairer, S.P. N{\o }rsett, and G.~Wanner.
\newblock {\em Solving Ordinary Differential Equations I}.
\newblock Springer-Verlag, Berlin, Germany, 1993.

\bibitem{Theocharis10}
G.~Theocharis, N.~Boechler, P.~G. Kevrekidis, S.~Job, M.~A. Porter, and
  C.~Daraio.
\newblock Intrinsic energy localization through discrete gap breathers in
  one-dimensional diatomic granular crystals.
\newblock {\em Phys. Rev. E}, 82:056604, 2010.

\bibitem{hooge12}
C.~Hoogeboom, Y.~Man, N.~Boechler, G.~Theocharis, P.~G. Kevrekidis, I.~G.
  Kevrekidis, and C.~Daraio.
\newblock Hysteresis loops and multi-stability: {F}rom periodic orbits to
  chaotic dynamics (and back) in diatomic granular crystals.
\newblock {\em Euro. Phys. Lett.}, 101:44003, 2013.

\bibitem{Flach05}
S.~Flach and A.~V. Gorbach.
\newblock Discrete breathers in {F}ermi--{P}asta--{U}lam lattices.
\newblock {\em Chaos}, 15:015112, 2005.

\bibitem{ODE1}
C.C. Chicone.
\newblock {\em Ordinary differential equations with applications}.
\newblock Springer, New York, 1999.

\bibitem{carpe}
F.~Fraternali, G.~Carpentieri, A.~Amendola, R.~E. Skelton, and V.~F.
  Nesterenko.
\newblock Multiscale tunability of solitary wave dynamics in tensegrity
  metamaterials.
\newblock {\em Applied Physics Letters}, 105(20):201903, 2014.

\bibitem{origami}
H.~Yasuda, C.~Chong, E.~G. Charalampidis, P.~G. Kevrekidis, and J.~Yang.
\newblock Formation of rarefaction waves in origami-based metamaterials.
\newblock {\em Phys. Rev. E}, 93:043004, 2016.

\bibitem{origami2}
H.~Yasuda, Y.~Miyazawa, E.~G. Charalampidis, C.~Chong, P.~G. Kevrekidis, and
  J.~Yang.
\newblock Origami-based impact mitigation via rarefaction solitary wave
  creation.
\newblock {\em Science Advances}, 5:eaau2835, 2019.

\bibitem{RarefactionNester}
E.~B. Herbold and V.~F. Nesterenko.
\newblock Propagation of rarefaction pulses in discrete materials with
  strain-softening behavior.
\newblock {\em Phys. Rev. Lett.}, 110:144101, 2013.

\bibitem{HEC_DSW}
H.~Kim, E.~Kim, C.~Chong, P.~G. Kevrekidis, and J.~Yang.
\newblock Demonstration of dispersive rarefaction shocks in hollow elliptical
  cylinder chains.
\newblock {\em Phys. Rev. Lett.}, 120:194101, 2018.

\bibitem{James2013}
G.~James, P.~G. Kevrekidis, and J.~Cuevas.
\newblock Breathers in oscillator chains with {H}ertzian interactions.
\newblock {\em Physica D}, 251:39, 2013.

\bibitem{Huang2}
G.~Huang and B.~Hu.
\newblock Asymmetric gap soliton modes in diatomic lattices with cubic and
  quartic nonlinearity.
\newblock {\em Phys. Rev. B}, 57:5746, 1998.

\bibitem{hooge11}
C.~Hoogeboom and P.G. Kevrekidis.
\newblock Breathers in periodic granular chains with multiple band gaps.
\newblock {\em Phys. Rev. E}, 86:061305, 2012.

\bibitem{Boechler2010}
N.~Boechler, G.~Theocharis, S.~Job, P.~G. Kevrekidis, M.~A. Porter, and
  C.~Daraio.
\newblock Discrete breathers in one-dimensional diatomic granular crystals.
\newblock {\em Phys. Rev. Lett.}, 104:244302, 2010.

\bibitem{Stathis}
E.~G. Charalampidis, F.~Li, C.~Chong, J.~Yang, and P.~G. Kevrekidis.
\newblock Time-periodic solutions of driven-damped trimer granular crystals.
\newblock {\em Math. Probl. in Eng.}, 2015:830978, 2015.

\bibitem{pdimer}
M.~M. Lee, E.~G. Charalampidis, S.~Xing, C.~Chong, and P.~G. Kevrekidis.
\newblock Breathers in lattices with alternating strain-hardening and
  strain-softening interactions.
\newblock {\em Phys. Rev. E}, 107:054208, May 2023.

\bibitem{Theocharis2009}
G.~Theocharis, M.~Kavousanakis, P.~G. Kevrekidis, C.~Daraio, M.~A. Porter, and
  I.~G. Kevrekidis.
\newblock Localized breathing modes in granular crystals with defects.
\newblock {\em Phys. Rev. E}, 80:066601, 2009.

\bibitem{Defect_Job}
St\'ephane Job, Francisco Santibanez, Franco Tapia, and Francisco Melo.
\newblock Wave localization in strongly nonlinear {Hertzian} chains with mass
  defect.
\newblock {\em Phys. Rev. E}, 80:025602, Aug 2009.

\bibitem{Defect_Man}
Y.~Man, N.~Boechler, G.~Theocharis, P.~G. Kevrekidis, and C.~Daraio.
\newblock Defect modes in one-dimensional granular crystals.
\newblock {\em Phys. Rev. E}, 85:037601, 2012.

\bibitem{tube}
H.W. Liepmann and A.~Roshko.
\newblock {\em Elements of Gasdynamics}.
\newblock Wiley, New York, 1957.

\bibitem{piston}
R.~Courant and K.~O. Friedrichs.
\newblock {\em Supersonic Flow and Shock Waves}.
\newblock Springer-Verlag, Berlin, 1948.

\bibitem{Smoller}
J.~Smoller.
\newblock {\em Shock Waves and Reaction--Diffusion Equations}.
\newblock Springer-Verlag, New York, 1983.

\bibitem{first_DSW}
D.~H. Tsai and C.~W. Beckett.
\newblock Shock wave propagation in cubic lattices.
\newblock {\em J. of Geophys. Res.}, 71(10):2601, 1966.

\bibitem{holian_atomistic_1995}
B.~L. Holian.
\newblock Atomistic computer simulations of shock waves.
\newblock {\em Shock Waves}, 5(3):149--157, 1995.

\bibitem{talcohen}
Jian Li, S~Chockalingam, and Tal Cohen.
\newblock Observation of ultraslow shock waves in a tunable magnetic lattice.
\newblock {\em Phys. Rev. Lett.}, 127:014302, Jun 2021.

\bibitem{KdVToda_limits}
Christopher Chong, Ari Geisler, Panayotis Kevrekidis, and Gino Biondini.
\newblock Integrable approximations of dispersive shock waves of the granular
  chain.
\newblock {\em arXiv:2402.08218}, 2024.

\bibitem{Hascoet2000}
E.~Hascoet and H.~J. Herrmann.
\newblock Shocks in non-loaded bead chains with impurities.
\newblock {\em Eur. Phys. J. B}, 14:183, 2000.

\bibitem{Herbold07}
E.~B. Herbold and V.~F. Nesterenko.
\newblock Solitary and shock waves in discrete strongly nonlinear double
  power-law materials.
\newblock {\em Appl. Phys. Lett.}, 90(26):261902, 2007.

\bibitem{Molinari2009}
A.~Molinari and C.~Daraio.
\newblock Stationary shocks in periodic highly nonlinear granular chains.
\newblock {\em Phys. Rev. E}, 80:056602, 2009.

\bibitem{shock_trans_granular}
E.~B. Herbold and V.~F. Nesterenko.
\newblock Shock wave structure in a strongly nonlinear lattice with viscous
  dissipation.
\newblock {\em Phys. Rev. E}, 75:021304, 2007.

\bibitem{fleischer2}
Shu Jia, Wenjie Wan, and Jason~W. Fleischer.
\newblock Dispersive shock waves in nonlinear arrays.
\newblock {\em Phys. Rev. Lett.}, 99:223901, Nov 2007.

\bibitem{Whitham74}
G.B. Whitham.
\newblock {\em Linear and Nonlinear Waves}.
\newblock Wiley, New York, 1974.

\bibitem{DSWreview}
G.A. El, M.A. Hoefer, and M.~Shearer.
\newblock Dispersive and diffusive-dispersive shock waves for nonconvex
  conservation laws.
\newblock {\em SIAM Review}, 59(1):3, 2017.

\bibitem{scholar}
M.~J. Ablowitz and M.~Hoefer.
\newblock Dispersive shock waves.
\newblock {\em Scholarpedia}, 4(11):5562, 2009.

\bibitem{Venakides99}
A.~M. Filip and S.~Venakides.
\newblock Existence and modulation of traveling waves in particles chains.
\newblock {\em Comm. Pure and Appl. Math.}, 52(6):693, 1999.

\bibitem{DHM06}
W.~Dreyer, M.~Herrmann, and A.~Mielke.
\newblock Micro-macro transition in the atomic chain via {W}hitham's modulation
  equation.
\newblock {\em Nonlinearity}, 19(2):471, 2005.

\bibitem{El2005}
G.~A. El.
\newblock {Resolution of a shock in hyperbolic systems modified by weak
  dispersion}.
\newblock {\em Chaos: An Interdisciplinary Journal of Nonlinear Science},
  15(3):037103, 10 2005.

\bibitem{Sprenger2024}
Patrick Sprenger, Christopher Chong, Emmanuel Okyere, Michael Herrmann,
  Panayotis Kevrekidis, and Mark Hoefer.
\newblock Dispersive hydrodynamics of a discrete conservation law.
\newblock {\em arXiv:2404.1675}, 2024.

\bibitem{CHONG2022}
Christopher Chong, Michael Herrmann, and P.G. Kevrekidis.
\newblock Dispersive shock waves in lattices: A dimension reduction approach.
\newblock {\em Physica D: Nonlinear Phenomena}, 442:133533, 2022.

\bibitem{NIST2010}
F.W. Olver, D.W. Lozier, R.F. Boisvert, and C.W. Clark.
\newblock {\em NIST Handbook of Mathematical Functions}.
\newblock Cambridge University Press, 2010.

\bibitem{GP73}
A.~V. {Gurevich} and L.~P. {Pitaevskii}.
\newblock {Nonstationary structure of a collisionless shock wave}.
\newblock {\em Zhurnal Eksperimentalnoi i Teoreticheskoi Fiziki}, 65:590--604,
  1973.

\bibitem{Kamchatnov}
A~M Kamchatnov.
\newblock {\em Nonlinear Periodic Waves and Their Modulations}.
\newblock World Scientific, 2000.

\bibitem{SIAP59p2162}
Yuji Kodama.
\newblock The whitham equations for optical communications: Mathematical theory
  of nrz.
\newblock {\em SIAM Journal on Applied Mathematics}, 59(6):2162--2192, 1999.

\bibitem{MielkePatz2017}
Mielke Alexander and Patz Carsten.
\newblock Uniform asymptotic expansions for the fundamental solution of
  infinite harmonic chains.
\newblock {\em Z. Anal. Anwend.}, 36:437--475, 2017.

\bibitem{blochkodama}
A.~M. Bloch and Y.~Kodama.
\newblock Dispersive regularization of the whitham equation for the toda
  lattice.
\newblock {\em SIAM Journal on Applied Mathematics}, 52(4):909--928, 1992.

\bibitem{Gino2023}
Gino Biondini, Christopher Chong, and Panayotis Kevrekidis.
\newblock On the whitham modulation equations for the toda lattice and the
  quantitative characterization of its dispersive shocks.
\newblock {\em ArXiv:2312.10755}, 2023.

\bibitem{rosenau1}
P.~Rosenau.
\newblock Dynamics of nonlinear mass-spring chains near the continuum limit.
\newblock {\em Phys. Lett. A}, 118:222, 1986.

\bibitem{rosenau2}
P.~Rosenau.
\newblock Hamiltonian dynamics of dense chains and lattices: {O}r how to
  correct the continuum.
\newblock {\em Phys. Lett. A}, 311:39, 2003.

\bibitem{suyang}
S.~Yang, G.~Biondini, C.~Chong, and P.~G. Kevrekidis.
\newblock Dispersive shock waves in granular crystals without precompression.
\newblock {\em in preparation}, 2024.

\bibitem{Martinez_disorder_2016}
Alejandro~J. Mart\'{\i}nez, P.~G. Kevrekidis, and Mason~A. Porter.
\newblock Superdiffusive transport and energy localization in disordered
  granular crystals.
\newblock {\em Phys. Rev. E}, 93:022902, Feb 2016.

\bibitem{Kim_disorder_2016}
E.~Kim, A.~Mart{\'i}nez, S.~Phenisee, P.~G. Kevrekidis, M.~A. Porter, and
  J.~Yang.
\newblock Direct measurement of superdiffusive energy transport and energy
  localization in disordered granular chains.
\newblock {\em Nature Communications}, 9:640, 2018.

\bibitem{JAMES201339}
G.~James, P.~G. Kevrekidis, and J.~Cuevas.
\newblock Breathers in oscillator chains with hertzian interactions.
\newblock {\em Physica D: Nonlinear Phenomena}, 251:39--59, 2013.

\bibitem{Liu2015}
Lifeng Liu, Guillaume James, Panayotis Kevrekidis, and Anna Vainchtein.
\newblock Strongly nonlinear waves in locally resonant granular chains.
\newblock {\em Nonlinearity}, 29(11):3496, 2016.

\bibitem{Liu2016}
L.~Liu, G.~James, P.~G. Kevrekidis, and A.~Vainchtein.
\newblock Breathers in a locally resonant granular chain with precompression.
\newblock {\em arXiv:1603.06033}, 2016.

\bibitem{faver_dimer}
Timothy~E. Faver and Hermen~Jan Hupkes.
\newblock Micropterons, nanopterons and solitary wave solutions to the diatomic
  fermi–pasta–ulam–tsingou problem.
\newblock {\em Partial Differential Equations in Applied Mathematics},
  4:100128, 2021.

\bibitem{yuli}
K.~Vorotnikov, Y.~Starosvetsky, G.~Theocharis, and P.G. Kevrekidis.
\newblock Wave propagation in a strongly nonlinear locally resonant granular
  crystal.
\newblock {\em Physica D: Nonlinear Phenomena}, 365:27--41, 2018.

\bibitem{Stefanov2015}
H.~Xu, P.~G. Kevrekidis, and A.~Stefanov.
\newblock Traveling waves and their tails in locally resonant granular systems.
\newblock {\em J. Phys. A: Math. Theor.}, 48:195204, 2015.

\bibitem{jkprl}
E.~Kim, F.~Li, C.~Chong, G.~Theocharis, J.~Yang, and P.~G. Kevrekidis.
\newblock Highly nonlinear wave propagation in elastic woodpile periodic
  structures.
\newblock {\em Phys. Rev. Lett.}, 114:118002, 2015.

\bibitem{VAINCHTEIN2022133252}
Anna Vainchtein.
\newblock Solitary waves in {FPU}-type lattices.
\newblock {\em Physica D: Nonlinear Phenomena}, 434:133252, 2022.

\bibitem{moleron}
M.~Moler\'on, A.~Leonard, and C.~Daraio.
\newblock Solitary waves in a chain of repelling magnets.
\newblock {\em Journal of Applied Physics}, 115(18):184901, 2014.

\bibitem{magBreathers}
M.~Moler\'on, C.~Chong, A.~J. Mart\'inez, M.~A. Porter, P.~G. Kevrekidis, and
  C.~Daraio.
\newblock Nonlinear excitations in magnetic lattices with long-range
  interactions.
\newblock {\em New Journal of Physics}, 21:063032, 2019.

\bibitem{magBreathers2D}
C.~Chong, Y.~Wang, D.~Marechal, E.~G. Charalampidis, Miguel Moler\'on,
  Alejandro~J. Mart\'inez, Mason~A. Porter, P.~G. Kevrekidis, and Chiara
  Daraio.
\newblock Nonlinear localized modes in two-dimensional hexagonally-packed
  magnetic lattices.
\newblock {\em New Journal of Physics}, 23:043008, 2021.

\bibitem{Kim23}
B.~L. Kim, C.~Chong, S.~Hajarolasvadi, Y.~Wang, and C.~Daraio.
\newblock Dynamics of time-modulated, nonlinear phononic lattices.
\newblock {\em Phys. Rev. E}, 107:034211, Mar 2023.

\bibitem{Kim24}
Christopher Chong, Brian Kim, Evelyn Wallace, and Chiara Daraio.
\newblock Modulation instability and wavenumber bandgap breathers in a time
  layered phononic lattice.
\newblock {\em Phys. Rev. Res.}, 6:023045, Apr 2024.

\bibitem{FlachLong}
S.~Flach.
\newblock Breathers on lattices with long range interaction.
\newblock {\em Phys. Rev. E}, 58:R4116--R4119, Oct 1998.

\bibitem{FractionalBook}
P.G. Kevrekidis and J.~Cuevas.
\newblock {\em Fractional Dispersive Models and Applications}.
\newblock Springer-Verlag, Heidelberg, 2024.

\bibitem{Molina21}
Mario~I. Molina.
\newblock Fractional nonlinear electrical lattice.
\newblock {\em Phys. Rev. E}, 104:024219, Aug 2021.

\bibitem{maciasbountis2022}
T~Bountis J~E Macias~Diaz.
\newblock An efficient dissipation-preserving numerical scheme to solve a
  caputo--riesz time-space-fractional nonlinear wave equation.
\newblock {\em Fractal/Fractional}, 6(9):500--525, 2022.

\bibitem{topphoto}
Tomoki Ozawa, Hannah~M. Price, Alberto Amo, Nathan Goldman, Mohammad Hafezi,
  Ling Lu, Mikael~C. Rechtsman, David Schuster, Jonathan Simon, Oded
  Zilberberg, and Iacopo Carusotto.
\newblock Topological photonics.
\newblock {\em Rev. Mod. Phys.}, 91:015006, Mar 2019.

\bibitem{topological}
Zhihao Lan, Menglin~L.N. Chen, Fei Gao, Shuang Zhang, and Wei~E.I. Sha.
\newblock A brief review of topological photonics in one, two, and three
  dimensions.
\newblock {\em Reviews in Physics}, 9:100076, 2022.

\bibitem{ni2023topological}
Xiang Ni, Simon Yves, Alex Krasnok, and Andrea Alu.
\newblock Topological metamaterials.
\newblock {\em Chemical Reviews}, 123(12):7585--7654, 2023.

\bibitem{zhang2018topological}
Xiujuan Zhang, Meng Xiao, Ying Cheng, Ming-Hui Lu, and Johan Christensen.
\newblock Topological sound.
\newblock {\em Communications Physics}, 1(1):97, 2018.

\bibitem{xue2022topological}
Haoran Xue, Yihao Yang, and Baile Zhang.
\newblock Topological acoustics.
\newblock {\em Nature Reviews Materials}, 7(12):974--990, 2022.

\bibitem{xin2020topological}
Li~Xin, Yu~Siyuan, Liu Harry, Lu~Minghui, and Chen Yanfeng.
\newblock Topological mechanical metamaterials: A brief review.
\newblock {\em Current Opinion in Solid State and Materials Science},
  24(5):100853, 2020.

\bibitem{ABLOWITZ2022133440}
Mark~J. Ablowitz and Justin~T. Cole.
\newblock Nonlinear optical waveguide lattices: Asymptotic analysis, solitons,
  and topological insulators.
\newblock {\em Physica D: Nonlinear Phenomena}, 440:133440, 2022.

\bibitem{leykam2013flat}
Daniel Leykam, Sergej Flach, Omri Bahat-Treidel, and Anton~S. Desyatnikov.
\newblock Flat band states: Disorder and nonlinearity.
\newblock {\em Phys. Rev. B}, 88:224203, December 2013.

\bibitem{leykam2018artificial}
Daniel Leykam, Alexei Andreanov, and Sergej Flach.
\newblock Artificial flat band systems: from lattice models to experiments.
\newblock {\em Adv. Phys.: X}, 3(1):1473052, 2018.

\bibitem{phononic2d}
Pragalv Karki and Jayson Paulose.
\newblock Non-singular and singular flat bands in tunable phononic
  metamaterials.
\newblock {\em Phys. Rev. Res.}, 5:023036, Apr 2023.

\bibitem{acoustic3d}
Ya-Xi Shen, Yu-Gui Peng, Pei-Chao Cao, Jensen Li, and Xue-Feng Zhu.
\newblock Observing localization and delocalization of the flat-band states in
  an acoustic cubic lattice.
\newblock {\em Phys. Rev. B}, 105:104102, Mar 2022.

\bibitem{FlatbandElectric}
Carys Chase-Mayoral, L.~Q. English, Noah Lape, Yeongjun Kim, Sanghoon Lee,
  Alexei Andreanov, Sergej Flach, and P.~G. Kevrekidis.
\newblock Compact localized states in electric circuit flat-band lattices.
\newblock {\em Phys. Rev. B}, 109:075430, Feb 2024.

\bibitem{kagomejoh}
Rodrigo~A. Vicencio and Magnus Johansson.
\newblock Discrete flat-band solitons in the kagome lattice.
\newblock {\em Phys. Rev. A}, 87:061803, Jun 2013.

\bibitem{mukherjee2015observation}
Sebabrata Mukherjee, Alexander Spracklen, Debaditya Choudhury, Nathan Goldman,
  Patrik \"Ohberg, Erika Andersson, and Robert~R. Thomson.
\newblock Observation of a localized flat-band state in a photonic lieb
  lattice.
\newblock {\em Phys. Rev. Lett.}, 114:245504, Jun 2015.

\bibitem{vicencio2015observation}
Rodrigo~A. Vicencio, Camilo Cantillano, Luis Morales-Inostroza, Basti\'an Real,
  Cristian Mej\'{\i}a-Cort\'es, Steffen Weimann, Alexander Szameit, and
  Mario~I. Molina.
\newblock Observation of localized states in {Lieb} photonic lattices.
\newblock {\em Phys. Rev. Lett.}, 114:245503, June 2015.

\bibitem{morales2016simple}
Luis Morales-Inostroza and Rodrigo~A. Vicencio.
\newblock Simple method to construct flat-band lattices.
\newblock {\em Phys. Rev. A}, 94:043831, October 2016.

\bibitem{andrea}
A.~Leonard, C.~Chong, P.~G. Kevrekidis, and C.~Daraio.
\newblock Traveling waves in {2D} hexagonal granular crystal lattices.
\newblock {\em Granular Matter}, 16(4):531, 2014.

\bibitem{ruzze}
Raj~Kumar Pal, Javier Vila, Michael Leamy, and Massimo Ruzzene.
\newblock Amplitude-dependent topological edge states in nonlinear phononic
  lattices.
\newblock {\em Phys. Rev. E}, 97:032209, Mar 2018.

\bibitem{DMD_topological}
Shuaifeng Li, Panayotis~G. Kevrekidis, and Jinkyu Yang.
\newblock Characterization of elastic topological states using dynamic mode
  decomposition.
\newblock {\em Phys. Rev. B}, 107:184308, May 2023.

\bibitem{DMD_topological2}
Joshua~R. Tempelman, Alexander~F. Vakakis, and Kathryn~H. Matlack.
\newblock A modal decomposition approach to topological wave propagation.
\newblock {\em Journal of Sound and Vibration}, 568:118033, 2024.

\bibitem{colbrook2023multiverse}
Matthew~J. Colbrook.
\newblock The multiverse of dynamic mode decomposition algorithms, 2023.

\bibitem{James_2021}
Guillaume James.
\newblock Traveling fronts in dissipative granular chains and nonlinear
  lattices.
\newblock {\em Nonlinearity}, 34(3):1758, mar 2021.

\bibitem{raissi_physics-informed_2019}
M.~Raissi, P.~Perdikaris, and G.~E. Karniadakis.
\newblock Physics-informed neural networks: {A} deep learning framework for
  solving forward and inverse problems involving nonlinear partial differential
  equations.
\newblock {\em Journal of Computational Physics}, 378:686--707, February 2019.

\bibitem{brunton_discovering_2016}
S.L. Brunton, J.L. Proctor, and J.N. Kutz.
\newblock Discovering governing equations from data by sparse identification of
  nonlinear dynamical systems.
\newblock {\em Proceedings of the National Academy of Sciences},
  113(15):3932--3937, April 2016.

\bibitem{lu2021deepxde}
L.~Lu, X.~Meng, Z.~Mao, and G.E. Karniadakis.
\newblock Deep{XDE}: A deep learning library for solving differential
  equations.
\newblock {\em SIAM Review}, 63(1):208--228, 2021.

\bibitem{karniadakis2021physics}
G.E. Karniadakis, I.G. Kevrekidis, L.~Lu, P.~Perdikaris, S.~Wang, and L.~Yang.
\newblock Physics-informed machine learning.
\newblock {\em Nature Reviews Physics}, 3(6):422--440, 2021.

\bibitem{SAQLAIN2023107498}
Sheikh Saqlain, Wei Zhu, Efstathios~G. Charalampidis, and Panayotis~G.
  Kevrekidis.
\newblock Discovering governing equations in discrete systems using pinns.
\newblock {\em Communications in Nonlinear Science and Numerical Simulation},
  126:107498, 2023.

\end{thebibliography}

\end{document}